\documentclass[aps,preprint]{revtex4}

\usepackage{epsfig,psfig,epsf,amsmath}

\def\bxi{\mbox{\boldmath{$\xi$}}}

\def\e{{e}}

\def\bu{{\bf u}}
\def\bv{{\bf v}}

\def\F{{\bf F}}

 \def\bJ{{\sf J}}
\def\bW{{\sf W}}  

\def\J{{\sf J}}
\def\W{{\sf W}}

\def\H{{\sf H}}
\def\i{{\rm i}}
\def\bH{{\sf H}} 

\def\cc{{\rm c.c.}}

\def\Re{{\sf Re}}    
\def\Im{{\sf Im}}    

\def\bc{\begin{center}}
\def\ec{\end{center}}
\def\be{\begin{equation}}
\def\ee{\end{equation}}
\def\bi{\begin{itemize}}
\def\ei{\end{itemize}}
\def\bea{\begin{eqnarray}}
\def\eea{\end{eqnarray}}

\begin{document}

\title{
Effects of noise in a cortical neural
model
}
\author{
Maria Marinaro
and Silvia Scarpetta
 }

\address   { Dipartimento di Fisica``E. R. Caianiello'', 
Salerno University, Baronissi (SA) IT \\
 INFM Sezione di Salerno (SA) IT, {\it and}
 IIASS Vietri sul mare (SA) Italy
} 
%

\begin{abstract}
Recently Segev et al. (Phys. Rev. E 64,2001, Phys.Rev.Let. 88, 2002)
 made long-term  observations
of spontaneous activity of in-vitro cortical networks,
which differ from predictions of current models in many features. 
In this paper we generalize the EI cortical model introduced
in a previous paper (S.Scarpetta et al. Neural Comput. 14, 2002),
including 
intrinsic white noise and analyzing
effects of  noise on the spontaneous activity of the nonlinear system,
in order to account for the experimental results of Segev et al..
Analytically we can distinguish different regimes of activity, depending
from the model parameters.
Using analytical results as a guide line, we perform simulations of the
nonlinear stochastic model in two different regimes, B and C. The Power 
Spectrum Density (PSD) of the activity and 
the Inter-Event-Interval (IEI) distributions are computed,
 and compared with experimental results.
In 
regime B 
the network shows 
stochastic resonance phenomena and  noise induces
aperiodic collective synchronous oscillations that mimic experimental observations
at 0.5 mM Ca concentration.  
In regime C
 the model shows spontaneous synchronous periodic activity that mimic
activity observed at 1 mM Ca concentration and the PSD 
shows two peaks at the $1^{th}$ and $2^{nd}$ harmonics  in agreement with
experiments at 1 mM Ca. Moreover (due to intrinsic noise and nonlinear
 activation function effects)
 the PSD shows a broad band peak at low 
frequency.
This feature, observed experimentally, does not find explanation 
in the previous models.
Besides we identify parametric changes 
(namely
increase of  noise or  decreasing of excitatory
connections)
that  
reproduces the fading of periodicity
found experimentally at long times, 
 and we identify a way to discriminate between
those two possible effects measuring experimentally the low frequency PSD. 
\end{abstract}

\maketitle



\section{ Introduction and motivations }


Spontaneous (and stimulation-driven)
 synchronized oscillatory activity has been
observed in many in vitro and in vivo experiments \cite{osc1,osc2,osc3}.

The understanding of spontaneous activity
 of in-vitro networks \cite{gross,pre02,segevl,segev} is a preliminary 
requirements for the
comprehension of the network behavior inside animal brain,
where the dynamics is more complex due to the presence of
external stimuli and  of interactions among different parts of the
brain.
In particular, the understanding of the
specific mechanisms underlying  spontaneous spatio-temporal
pattern of activity is important for the comprehension of
brain activity, especially in relation
with  epilepsy \cite{epilepsy}, the central pattern generator systems,
etc.

Recently Segev et al. \cite{segevl,segev}  have done accurate long-term
measurements of  the spontaneous activity
of  {\it in-vitro} 
 cortical cells neural networks placed on multi-electrode
arrays.
 The effect of external Ca concentration on the spontaneous activity has 
 been  studied.
They observed, for a critical range of Ca concentration,
periodic synchronized bursting activity,
that fades away after about $\sim 20 $ min.       
Periodic  synchronized bursting  is observed at $1 mM$ Ca concentration
and not at  higher (2 mM) and lower (0.5 mM) concentrations.
Their observations differ from prediction of current neural network models
in many features.
They try to model the phenomena performing numerical simulations of
an integrate-and-fire network model with random connections. 
Adding (1) dynamic threshold and (2) activity-dependent synaptic
connections, they reproduce much of the observed network activity,
but do not obtain a complete explanation of the
experimental results.
For example,  their numerical simulations capture 
the transition 
between aperiodic synchronized bursting versus periodic synchronized bursting
when  Ca concentration 
(and so model connection strengths) was increased,
but they do not reproduce the transition from
periodic synchronized bursting versus aperiodic activity 
observed experimentally when Ca concentration was increased
over the critical interval. 
Moreover the experimental data show that 
the  energy distribution over low frequencies
has a broad band with power law decay  that indicates the existence of
positive long range time correlations in the sequences of bursts, this
behavior 
cannot be accounted for by the Segev et al. model. 
The IEI distribution at 0.5 mM Ca shows a very long tail (tens of secs), 
 decaying much more slowly that in the  IF model. 
Finally, experimental results shows that,  after
 1mM Ca concentration was obtained, 
the high peaks of the PSD,
at first and second harmonic,
become lower and lower with time, 
and after $\sim 20$ min the PSD is almost flat.
These PSD features (in particular the behavior at low frequency and
 the changes
 that happen in time scale of several minutes)
have not been explained by the model of Segev et al 
\cite{segevl,segev}, and as far as we know, until now the
explanation is lacking.
%
Patterns of
spontaneous activity in cortical cultured networks were observed
 by other researchers \cite{osc3,pre02}
Canepari et. al \cite{osc3} observed 
transitions from asynchronous firing dynamics to synchronous firing
dynamics when Ca2+ concentration was increased from 0.1. mM
to 1 mM.
%
%

Here we  model a cultured cortical cells  network
 using a Excitatory-Inhibitory (EI) Cowan-Wilson-like \cite{cowan}
model analytically tractable,
which enable us to study the source of spontaneous activity dynamics
analytically.
We analyze the macroscopic observable that
describe the dynamics of the system
 as a function of the network parameters,
exploiting  effects of noise and nonlinearity.

We show that some insights and 
a good agreement with experiments 
can be obtained  including intrinsic white noise
in the  spiking-rate Excitatory-Inhibitory (EI)  neural 
network model.
The model is based on the noise-less cortical model introduced in a previous paper 
\cite{SZJ} 
by one of us,  Z. Li and J. Hertz, which is 
 able to imprint and retrieve
 oscillatory patterns when driven by noise-less oscillatory input (using 
a generalized hebbian learning rule).
%
Here the model is studied to put in evidence its
 spontaneous activity
and specifically  the effects of noise on the dynamics are analyzed.
The spatio-temporal patterns of spontaneous  activity 
in our model is a consequence
of the  dynamics
of the interacting  excitatory and inhibitory units,
 and depends critically
from the presence of noise
 and from the synaptic strengths of the EI network.

The results we have found, although referred to the 
work of Segev et al.\cite{segevl,segev}, are of more general interest,
indeed they connect the behavior of in-vitro neural networks to the
one of nonlinear subthreshold and overthreshold systems in presence of noise.

The effects of noise in neural models is the focus of a lot of recent
literature \cite{Shim02,pikovsky97,Gerstner00}, 
 from the single FitzHugh-Nagumo system point
 of view \cite{pikovsky97} to the  IF homogeneous networks
\cite{Gerstner00} or the phase-model coupled oscillators \cite{Shim02}. 
%

In section \ref{secII} the model is described, in sec \ref{secIII}
analytical and numerical results are reported and in section \ref{conclusion}
 there are discussions and conclusions.
\section{The model}
\label{secII}
Our starting point in modeling are the 
stochastic Cowan-Wilson-like EI equations \cite{cowan,Peter_book,SZJ}
 governing 
the state variables, modeling the membrane potentials,
$\bu=\{u_1, \ldots,$ $ u_N\}$ and $\bv=\{v_1, \ldots, v_N\}$,
respectively for the  excitatory and  inhibitory units:
\bea
\dot{u}_i&=& -\alpha u_i - \sum_j H_{ij} g_v(v_j) +
 \sum_{j} J_{ij} g_u(u_j) +
  \bar F_{i}(t)  ,						\label{eqnsu}		\\
\dot{v}_i&=& -\alpha v_i + 
 \sum_{j } W_{ij} g_u(u_j) +   F_{i} (t) .
		\label{eqnsv} 
\eea
The unit outputs $ g_u(u_1), \ldots, g_u(u_N)$ and $g_v(v_1), \ldots,$  $
g_v(v_N)$ represent the probabilities of the
cells firing (or instantaneous firing rates) 
where
$g_u$
$g_v$ are sigmoidal activation functions that model the neuronal
input-output
relations.
  $\alpha^{-1}$ is a time constant
 (about few milliseconds,
 for simplicity it is assumed equal for excitatory and inhibitory units)
modeling the  membrane time constant,
$ J_{ij}$ is the synaptic connection strength
from excitatory unit $j$ to excitatory unit $i$, $ W_{ij}$ is the
synaptic connection
strength from excitatory unit $j$ to inhibitory unit $i$
and $ H_{ij}$ is the
synaptic connection strength from inhibitory unit $j$ to excitatory unit
$i$.
Since in cortical area pyramidal cells have long range connections 
to others pyramidal cells and to inhibitory interneurons,
 while inhibitory
interneurons generally only project locally, we assume $ J_{ij}$ and $ W_{ij}$
to be long range connections, and $ H_{ij}$ to be local.
All these parameters are non-negative; the inhibitory character of the second
term on the right-hand side of Eqn.\ (\ref{eqnsu}) is indicated by the minus
sign preceding it.
$\bar F_{ i} (t)$ and $F_{ i} (t)$ model the intrinsic noise,
respectively  on the 
excitatory and inhibitory units,
 not included in the definition of $u,v$.
In a (cultured) interacting neurons system
  noise can be due to several reasons, like
thermal fluctuation,  ionic channel 
stochastic activities, and many others.
%
 We take the  noise
$\bar F_{ i} (t), F_{ i} (t)$ 
 to be
uncorrelated white noise,
such that $\langle ~F_i(t)~\rangle =  \langle ~\bar
F_i(t)~\rangle = 0$
  and $\langle~F_i(t)~F_j(t')~\rangle = \Gamma \delta_{ij} \delta_{t-t'}$,
$ \langle~\bar F_i(t)~\bar F_j(t')~\rangle =~\bar\Gamma \delta_{ij}
\delta_{t-t'}$.  
Each such unit $u_i, v_i$ 
 represents a local assembly of pyramidal cells or local
interneurons sharing common, or at least highly correlated, input. 
(The
number of neurons represented by the excitatory units may be in general
different from the number represented by the inhibitory units.) 
 For reasons explained in the
following, we choose the connection matrices $\J,\H, \W$ 
symmetric, apart from small random fluctuations.
This symmetry does not imply symmetry of the total 
connection matrix $ {  \left(
\begin{array}{cc} 
 \J & -\H \\ \W & 0\\ 
\end{array} 
\right)
} $, indeed it is highly asymmetric, as well
as the connections between excitatory and
inhibitory neurons are still asymmetric in this scenario.
%
%
%
%
%
%
%
%
%
\subsection{Network connectivity}

Since the connectivity formation  of the cultured system we want to model
has grown randomly and spontaneously,
we can reasonably assume that strength of each connection
is a function only of
the type of  pre-synaptic and post-synaptic neurons (Excitatory or
Inhibitory), and  of the
 distance  between them, plus eventually some random quenched fluctuations.
Recent estimation of connectivity in in-vitro rat cortical networks
has shown long range connections, with arborization of the neurons
of $1.2\pm 0.5 mm^2$ \cite{pre02}, such that each
 neuron was connected with about  600 nearby neurons.

We will analyze two type of connectivity structures.
\bi
\item In the first case, each excitatory unit is connected to all
the other units of the model, while inhibitory units only project locally.
In particular, J and W are long-range matrices 
given by 
\bea 
J_{ij} &=& j_0 ( 1+ \epsilon \eta^{(J)}_{ij} )/N \nonumber \\
 W_{ij} &=& W_0 ( 1 + \epsilon \eta^{(W)}_{ij})/N  \label{infinite}
\eea
and the matrix $H$ is local
$H_{ij}= h_0 \delta_{ij} ( 1 +  \epsilon \eta^{(H)}_{i})$,
 where $\epsilon\ll 1$ and $\eta^{(J)}_{ij}$, $\eta^{(W)}_{ij}$,
$\eta^{(H)}_{i}$
are random quenched values, uniformly distributed between -1 and 1.
When $\epsilon=0$, the three connection matrices commute each other
and share a complete set of eigenvectors. In particular the principal
eigenvector $\bxi_0 =( 1/\sqrt{N}, \dots, 1/\sqrt{N})$ has eigenvalues
$j_0$, $W_0$ and $h_0$, while the others $N-1$ eigenvectors
have eigenvalues 0, 0 and $h_0$, respectively. 
When $\epsilon>0$, the vector
$\bxi_0=( 1/\sqrt{N}, \dots, 1/\sqrt{N})$ will still be 
an eigenvector of the connection matrices, apart from 
corrections of order $O(\epsilon/\sqrt{N})$.
In this paper we take $\epsilon=0$,
in a following paper we will investigate the effects of order $\epsilon$
and $\epsilon^2$ (numerical simulations with small $\epsilon\neq 0$
give results qualitatively similar to the ones with $\epsilon=0$).

\item
In the second case, we consider structured short-range connectivity,
inspired by recent measurements \cite{segev3,pre02}.
We put one excitatory and one inhibitory unit on each site of a
square lattice with L rows and M columns, so that there will be N=LM
excitatory units and N inhibitory units.
Each inhibitory unit is connected only locally,
 to the excitatory unit that is on
the same site.
Numbering the unit in a type-writer way, from 0 to N-1, 
we connect the i-th excitatory unit to the excitatory and inhibitory units
that are on the eight sites numbered $i-1 \mod N, i+1 \mod N,
i-M \mod N, i+M \mod N,
i-M+1 \mod N, i+M-1\mod N,
i+M+1 \mod N,i-M-1 \mod N.$

Therefore, each elements of $J_{ij}$ is
\be
  J_{ij}=  \left\{ \begin{array}{cc}
           j_0/8  & \mbox{ if}  |i-j| \mod N =  1 \\
           j_0/8  & \mbox{ if}  |i-j| \mod N = M \\
           j_0/8  & \mbox{ if}  |i-j| \mod N = M \pm 1 \\
           0  & \mbox{ otherwise} \\
\end{array} \right.
\label{finite}
\ee
Analogously
 $W_{ij}=w_0/8 $ if $|i-j| \mod N = 1
\, \mbox{ or }\, M \mbox{ or }\, M\pm 1 $, and is zero otherwise.
$H_{ij}$ is $h_0 $ if $ i=j $ and is zero otherwise.

Notice that, apart from the units on the boundary, each unit is connected to
its four nearest neighbors and to its four next-nearest neighbors.
Furthermore the connection matrices J and W are Toplitz matrices
 so that
eigenvectors are the Fourier basis 
\be
\xi_n(j)=\frac{1}{\sqrt{N}} e^{i 2 \pi n
j/N  }
\label{eigenvector}
\ee 
and eigenvalues, given the notation $J_{ij} = J(i-j)=J(x)$,
can be easily calculated:  $j_n= 
\sum_{x=0}^{N-1} J(x) cos( \frac{2 \pi}{N} n x)  $,
and analogously for $\bW$.
The highest eigenvalue of $\bJ$ is $j_0$ (and $w_0$ for $\bW$), corresponding
to the eigenvector with n=0.
H is diagonal and have all eigenvalues equal to $h_0$.
We investigate also the case of structured short-range connectivity
 with open boundary conditions. In such a case
the eigenvectors and eigenvalues are not known analytically
and have to be computed numerically.

In experiments of Segev et al. \cite{segevl}   3 networks are analyzed with
different geometries and size: 
a small 50-cells with a quasi-1D 2mm x 50 $\mu m$ geometry,
medium $10^4$-cells networks, with a rectangular 2mm x 2 mm geometry,
and a large 2 $10^6$-cells network with a circular 11-mm-radius geometry.

When L=M our lattice model describe the 2D square geometry,
while when M=1 and L=N there are only 2 connections for each site
(left and right next neighbor sites)
and the model describe the quasi-1D geometry used in the
experiments.

In both cases 
(\ref{infinite}) and (\ref{finite}), all the matrices $\bJ,\bW$ and $\bH$ commute each 
other and share a
common set of eigenvectors given by the Fourier basis in eq.
(\ref{eigenvector}).

\ei

In the next section the model dynamics is analyzed in terms of
the eigenvectors and eigenvalues of the connection matrices.
The procedure is the same 
for both the structured short range connectivity (\ref{finite}) and
the long-range connectivity that we have considered,
since in both cases we know analytically the eigenvectors and eigenvalues of
the matrices.
In the open boundary conditions case we compute the eigenvalues numerically.

\section{Model dynamics 
}
\label{secIII}

Using vectorial notation, the dynamics of the model is described by the 2N
components vector $\{\bu,  \bv\}$ 
Let's call $\{\bar\bu, \bar \bv\}$  the fixed point 
determined by $\dot{\bu}=0,\dot{\bv}=0$ with  $\bar
\F(t)=0,\F(t)=0$.


Linearizing the  equations 
(\ref{eqnsu},\ref{eqnsv}) around the fixed point $\{\bar\bu, \bar \bv\}$,
	eliminating
$\bv$ from the equations,
and 
assuming noise to be only on the $\bv$ units ($\bar \Gamma= 0$),
 we get
\begin{equation}
\ddot{\bu} + (2\alpha -{\sf J}) \dot{\bu} + [\alpha^2 -
\alpha {\sf J} + {\sf H} {\sf W})] \bu = - \H \F(t) 
                        \label{ddu}
\end{equation}     
where $\bu$ is now measured from the fixed point value $\bar \bu$, 
and nonlinearity enters only
through the redefinition of the elements of $\bJ$ $\bH$ and $\bW$:
 $J_{ij} g'_u(\bar u_j) \rightarrow J_{ij}$,
$H_{ij} g'_u(\bar v_j) \rightarrow  H_{ij}$
 and $W_{ij} g'_u(\bar u_j) \rightarrow W_{ij}$.
We use bold and sans serif notation (e.g. $\bu,\bJ$) for vectors and
matrices respectively.

The fixed point $(\bar{\bu},\bar{\bv})$ 
is stable if the homogeneous associate equation of the Eqn.\ (\ref {ddu}) 
has only decaying solutions ${\bu}$.
In particular, when ${\sf J}, {\sf H}$
 and ${\sf W}$  share the same set of
eigenvectors ${\bxi_n}$,
denoting  with 
$ { j_n},  { W_n}$ and
 $ { h_n} $
their eigenvalues,
the eigensolutions are 
\be
{\bu_n} = e^{\lambda_n t} {\bxi_n},
\ee
where
\begin{equation}
\lambda_n= - \frac{ 2\alpha -{ {j_n} }}{2} 
\pm  \frac{ \sqrt{  {j_n}^2  - 4 h_n W_n } }{2} 
\label{la}
\end{equation}    
and the  stability condition is $\Re[\lambda_n]<0$
i.e., the real parts $\Re [ \lambda_n ] $  
 of the eigenvalues $\lambda_n$ 
of the homogeneous system must be negative.
The state vector $\bu$ is a linear combination of all the eigenmodes,
given by $\bu= \sum_n c_n  e^{\lambda_n t} {\bxi_n} + \cc$
in the absence of noise $\Gamma=0$,
therefore if all  the n modes have $\lambda_n= \Re[\lambda_n]<0$, 
and $\Im[\lambda_n]=0 $,
%
the activity $\bu$ simply decays toward
 the fixed point, 
stationary  state (regime A). 
In regime A
 the isolated system is "quiet" and, each unit
 just fire randomly, 
 each one uncorrelated with the others.
Regime A corresponds to $j_n< 2\alpha$ and $j_n^2 \simeq 4 h_n W_n$.
Besides the regime A,
others interesting dynamic regimes can be considered.
In the following we analyze two cases named
 respectively regime B and regime C.
The regime B arises when 
$\begin{array}{cc} 
 \Re[\lambda_n]<0&   \forall  n  \\
\Im[\lambda_n] = 0 &  \forall  n 
\,\,  \mbox{but one. \, (\small call it n=0)} \\
\end{array} $

It means that excitatory connections are such that
$j_0< 2\alpha $ and $j_0^2 < 4 h_0 W_0$.
In this regime, in absence of noise the system (after a transient 
with damped oscillations) settles down to
the stable fixed point. However
 as we will see later,  spontaneous collective aperiodic 
 oscillations are induced by
noise. The activity in presence of noise is similar to the one 
observed in-vitro by Segev et al. at 0,5 mM Ca concentration.
%
The regime C is present when it exists at least 
an eigenvalue, let call it $\lambda_0$, such that
$\begin{array}{c}
\Re[\lambda_0]>0 \\
 \Im[\lambda_0]\neq 0 \\
\end{array}$.

It means that $2\alpha<j_0<\sqrt{ 4 h_0 W_0 } $.
Therefore regime C occur only in a critical interval of
excitatory connections strength,
such that excitatory principal eigenvalue $j_0$ is greater
 then $2\alpha$ but
lower then $\sqrt{ 4 h_0 W_0 }$.
In this regime  
spontaneous oscillations grows in the linear approximation. 
As one expect, the saturating
nonlinearity stabilizes the limit cycle.
The nonlinear model shows 
synchronous periodic activity spontaneously, as we will
see later, similar to the experimental results 
observed at 1 mM Ca concentration.

When $j_0$ is so large that $j_0> 2\alpha$ and
 $j_0 > \sqrt{ 4 h_0 W_0 } $  the linear analysis predicts a divergence
without oscillations. In this case strong nonlinear effects have to taken in
consideration.  It has been observed by numerical simulations (sec.\ref{secC}) 
that the system shifts to a new stable fixed point, around which oscillations
are induced by noise.

The extracellular Ca2+ is known to affect  synapse probability of 
transmitter release \cite{14,17}
and  the action potential firing threshold (see \cite{15}).
The parameter of interest $\J, \H$ and $\W$
 are actually products of connection strengths and gradients of the
non-linear terms evaluated at the equilibrium position; calcium
concentration
may affect both terms.
So the effects of extracellular Ca2+ concentration may be multiple.
We model the increase of Ca concentration as an increase of
the excitatory connection strengths. 
Modeling the increase of Ca concentration as an increase
of excitatory connections  $\J$ and $\W$
(or at least as an increase of the excitatory-to-excitatory 
connection strength  $\J $),
then both the transitions observed experimentally  from 0.5 mM to 1 mM and 
from 1mM to 2 mM Ca concentration could be accounted by the model.
Indeed while a small increase in Ca concentration (and therefore a small
increase of $j_0$ w.r.t. $\alpha$)
induces a transition in the model
from regime B to regime C, a larger increase in Ca concentration
(and therefore a larger increase of $j_0$) make the model
to go out of regime C.


\section{Regime B Dynamics and Stochastic Resonance Effects }  
Starting from Eqn. (\ref{ddu}) we first analyze
the self-correlation function 
of excitatory unit i, 
$ C_i(t-t')=
 \langle u_i(t) u_i(t') \rangle   -  
\langle u_i(t)\rangle   \langle
u_i(t) \rangle
$
and
the average self-correlation function  
 $ C(t-t') = {\frac{1}{N} } \sum_i C_i(t-t')$ 
in the regime B in presence of noise of 
amplitude $\Gamma$.
%
The PSD,
i.e. the Fourier Transform  $\tilde C(\omega)$ of $C(t-t')$,
is the sum  of N contributions given by:
\be
\tilde C^n(\omega) \propto \frac{ h_0^2 \Gamma  \tau_n^4}
{ (1 + \omega^2 \tau_n^2)^2} 
\label{zero}
\end{equation}
when the eigenvalues are real and negative
 $\lambda_n=- 1/\tau_n $,
and by
\be
\!
\!\tilde C^0(\omega)\! \propto \!\frac{ h_0^2 \Gamma \tau_0^4}{4(1+\omega_0^2 \tau_0^2)}
[\! \frac{2+ \omega/\omega_0}{ 1+ \tau_0^2 (\omega+\omega_0)^2} 
\!+\! \frac{2 - \omega/\omega_0}{ 1 + \tau_0^2 (\omega-\omega_0)^2 }]
\label{pred_peak_w}
\end{equation}
for the eigenvalue $\lambda_0= - 1/\tau_0 + i \omega_0$,
with $\tau_0>0, \omega_0\neq 0$.
The contribution expressed by Eqn. (\ref{zero})
is peaked at $\omega=0$, while
the contribution given by Eqn. (\ref{pred_peak_w})
is peaked at $\omega$ close to $\omega_0$ (when $\tau_0 \omega_0 <1$).
In regime B the PSD has contributions also from Eqn. (\ref{pred_peak_w}).
Therefore the linear analysis predicts 
that  noise induces in the regime B a  collective 
 oscillatory behavior 
in the neurons activity.
This collective oscillatory behavior   
correspond to a broad peak in the power-spectrum
of the neurons activity near the characteristic 
frequency $\omega_0$.

\begin{figure}[ht]
\begin{center} \leavevmode
\begin{tabular}{ccc}
a&b&c
\vspace{-0.4cm}
\\
\hspace{-1.0cm}
\vspace{-0.4cm}
\epsfysize = 4.3cm 
\epsfbox{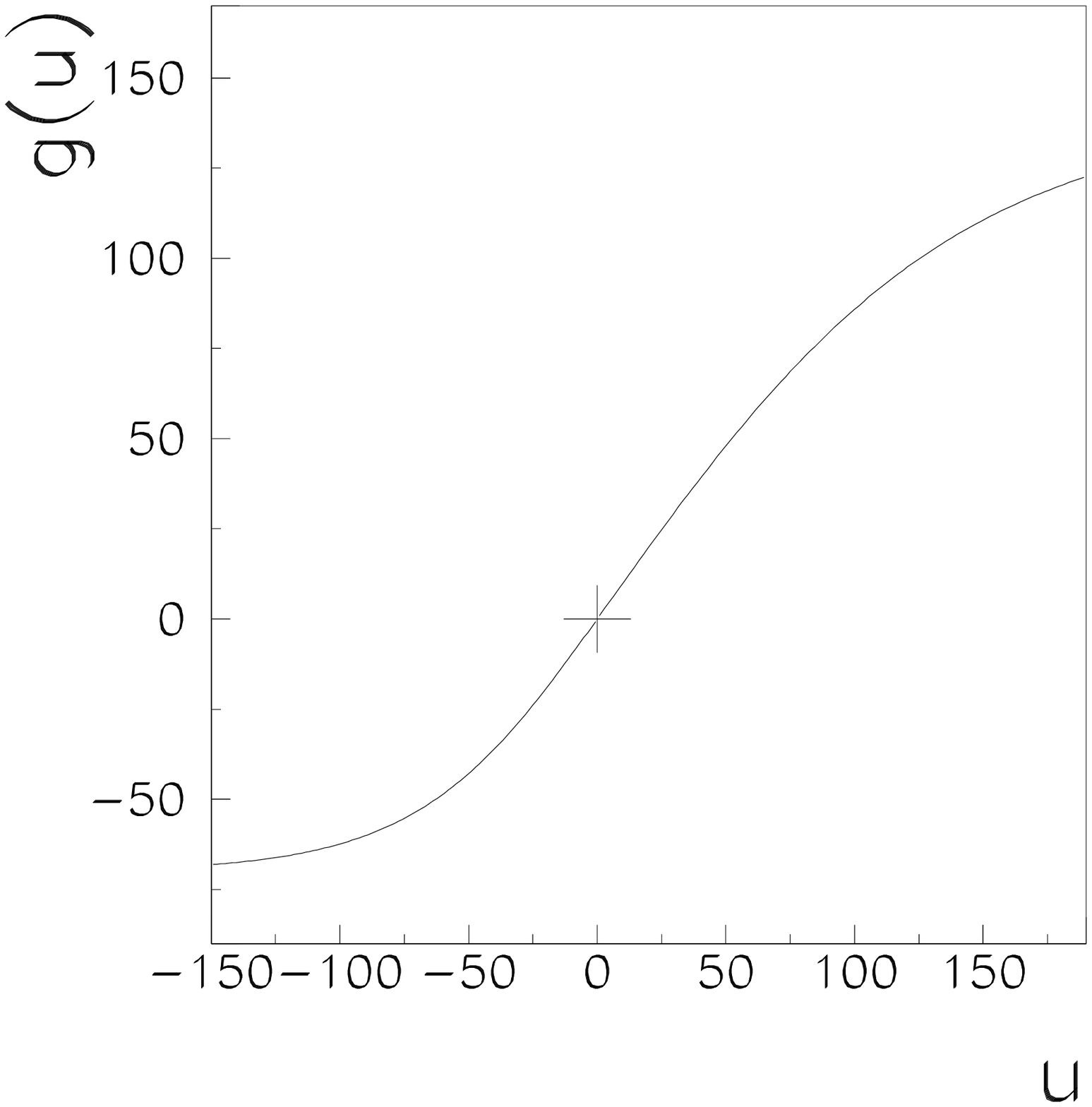}
&
\hspace{-0.6cm}
\vspace{-0.42cm} 
\epsfysize = 4.3cm  
\epsfbox{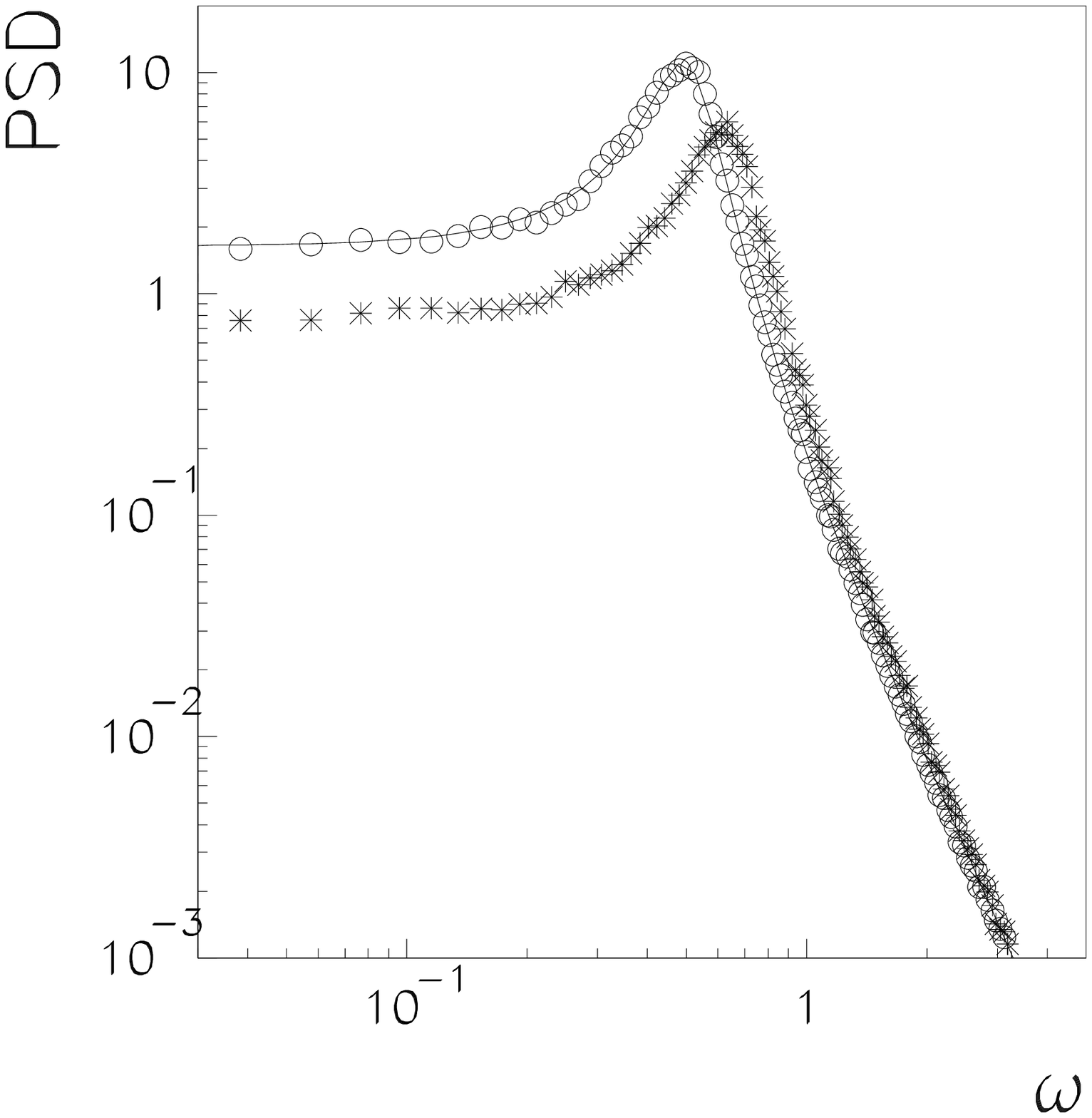}     
&
\hspace{-0.6cm}
\vspace{-0.42cm} 
 \epsfysize = 4.3cm
\epsfbox{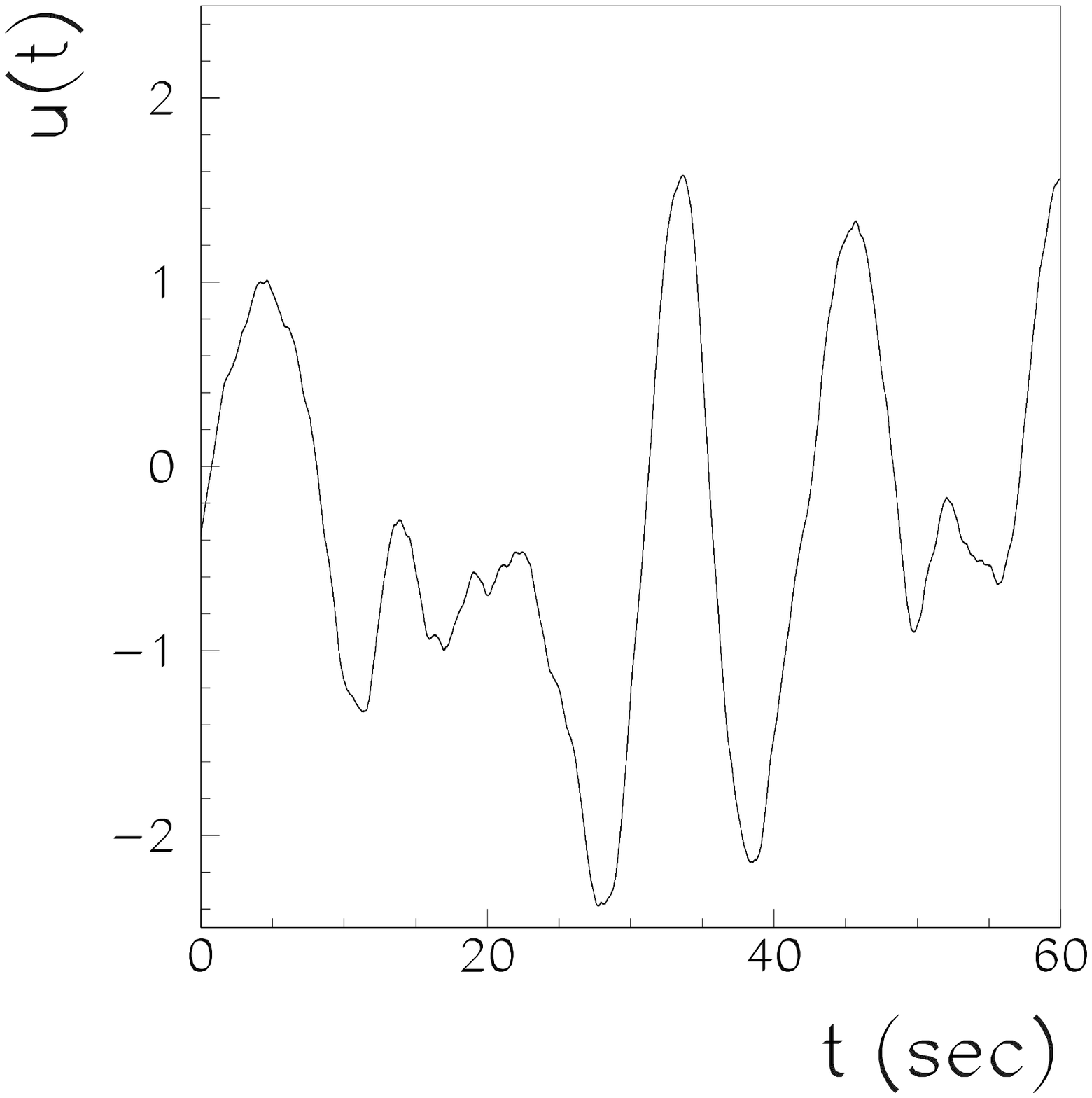} 
\end{tabular}
\end{center}
\vspace*{0.3cm}
\caption{
Numerical simulations of an long-range connectivity 
(eq.\ref{infinite}) model.
The values of parameters has been chosen in such a way to satisfy
the conditions for the regime B.
(specifically,  $\alpha= 50 sec^{-1}$, $N=10$,
$J_{ij}= \frac{j_0}{N} = \frac{ 2(\alpha -0.1)}{N} $,
$ W_{ij}= W_0/N ,  H_{ij}= h_0 \delta(i-j)$,
 $W_0=h_0= \sqrt{0.25 j_0^2+ 0.25} $,
so that $\omega_0= 
\sqrt{ - j_0^2/4 + h_0 W_0 } = 0.5 rad/sec$).
a. Activation function $g(u)$
used in simulations for excitatory units.
$u$ is measured from the fixed point value $\bar u$,
and $g(u)$ is shifted vertically so that the fixed point
coincides with the origin (0,0), marked by a cross.
b. Power Spectrum Density of  excitatory units activity,
 in  regime B, with noise $\Gamma=0.0004$.
Stars are nonlinear  simulations results,
circles are linear simulations results,
while  solid line is  theoretical prediction in linear
approximation.
A broad peak at $\omega\neq 0$ in the PSD is induced by noise.
c.
The time behavior of the state variable $u_i(t), i=1,\dots,N$
 in the linear numerical
simulation 
 in  regime B, with $\Gamma=0.0004$.
Lines $u_i(t)$, $i=1,..,N=10$,  overlaps each other because of synchrony.
All units $u_i(t)$ shows synchronous aperiodic oscillatory activity.
}
\label{fig_power_linB}
\end{figure}

\begin{figure}[ht] 
\begin{tabular}{cc}
a & b \\
\epsfysize = 5cm
\epsfbox{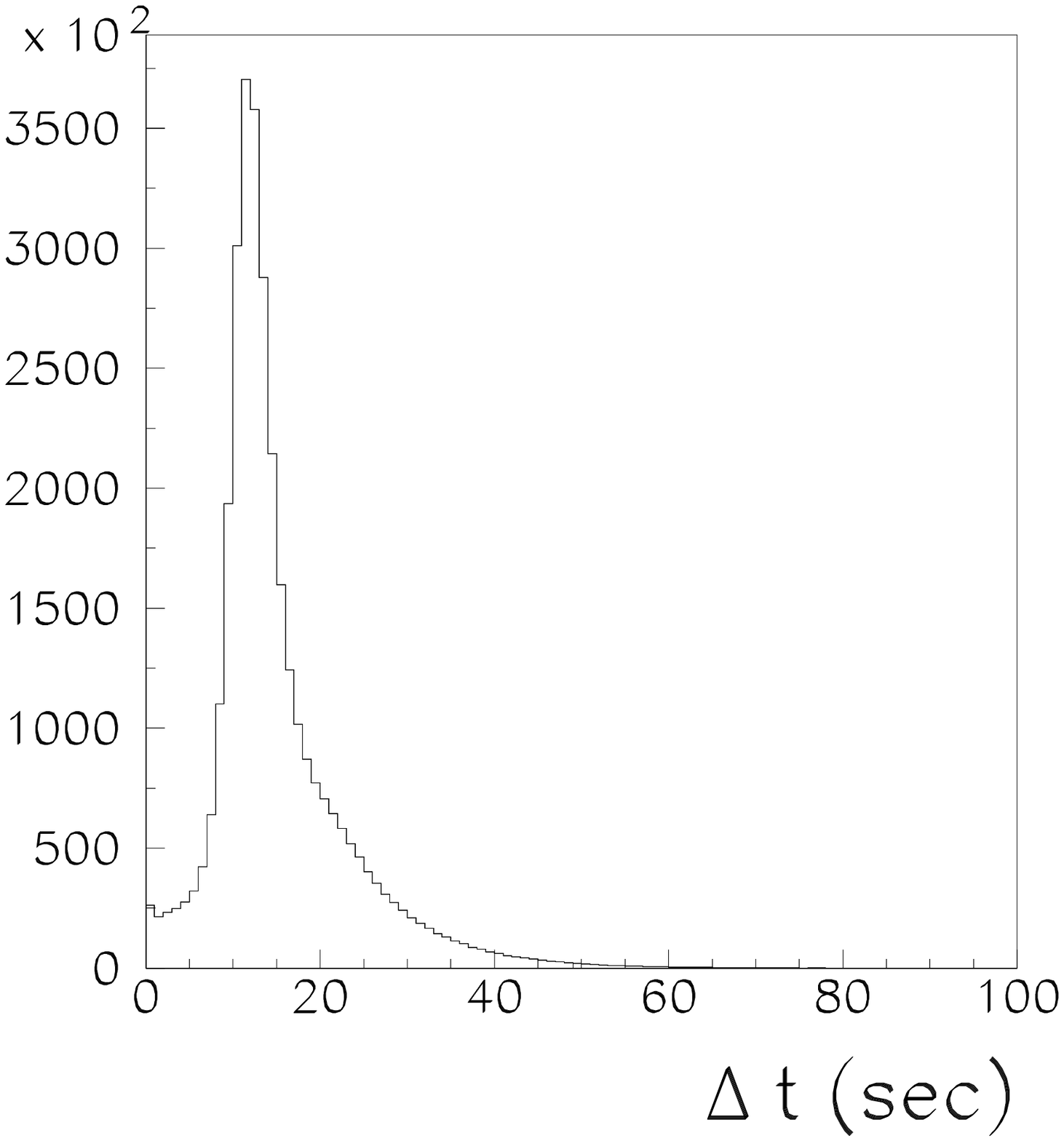} 
&
\epsfysize = 5cm
\epsfbox{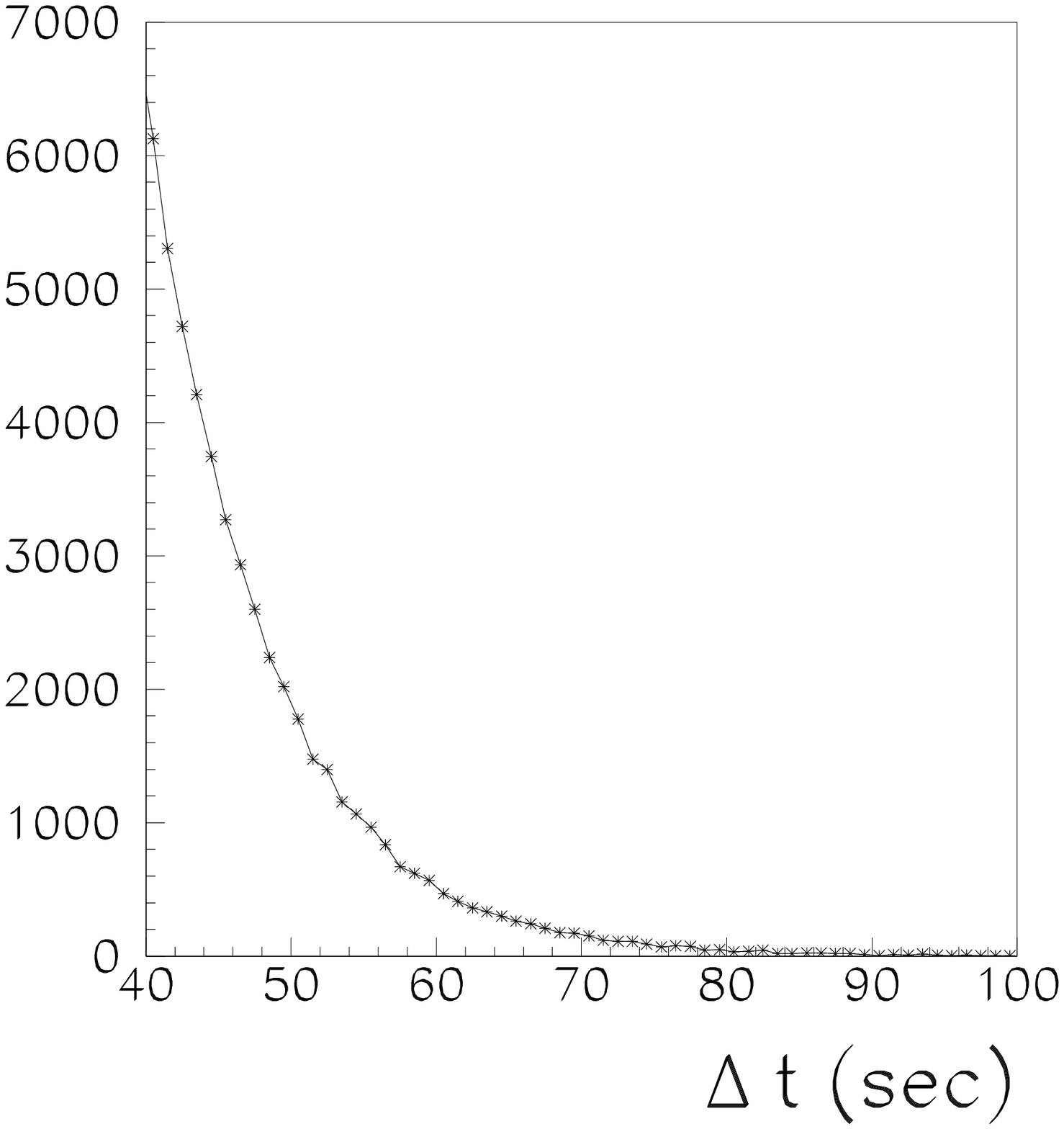}
\end{tabular}
\caption{a. Histogram of Inter Synchronous Event Intervals 
of the activity shows in previous figure (regime B with 
    $\Gamma=0.0004$).
    b. Zoom of the decay region of the IEI histogram.}
\label{figIEI}
\end{figure}

\begin{figure}[ht] 
\begin{tabular}{ccc}
a & b & c\\
\epsfysize = 5cm
\epsfbox{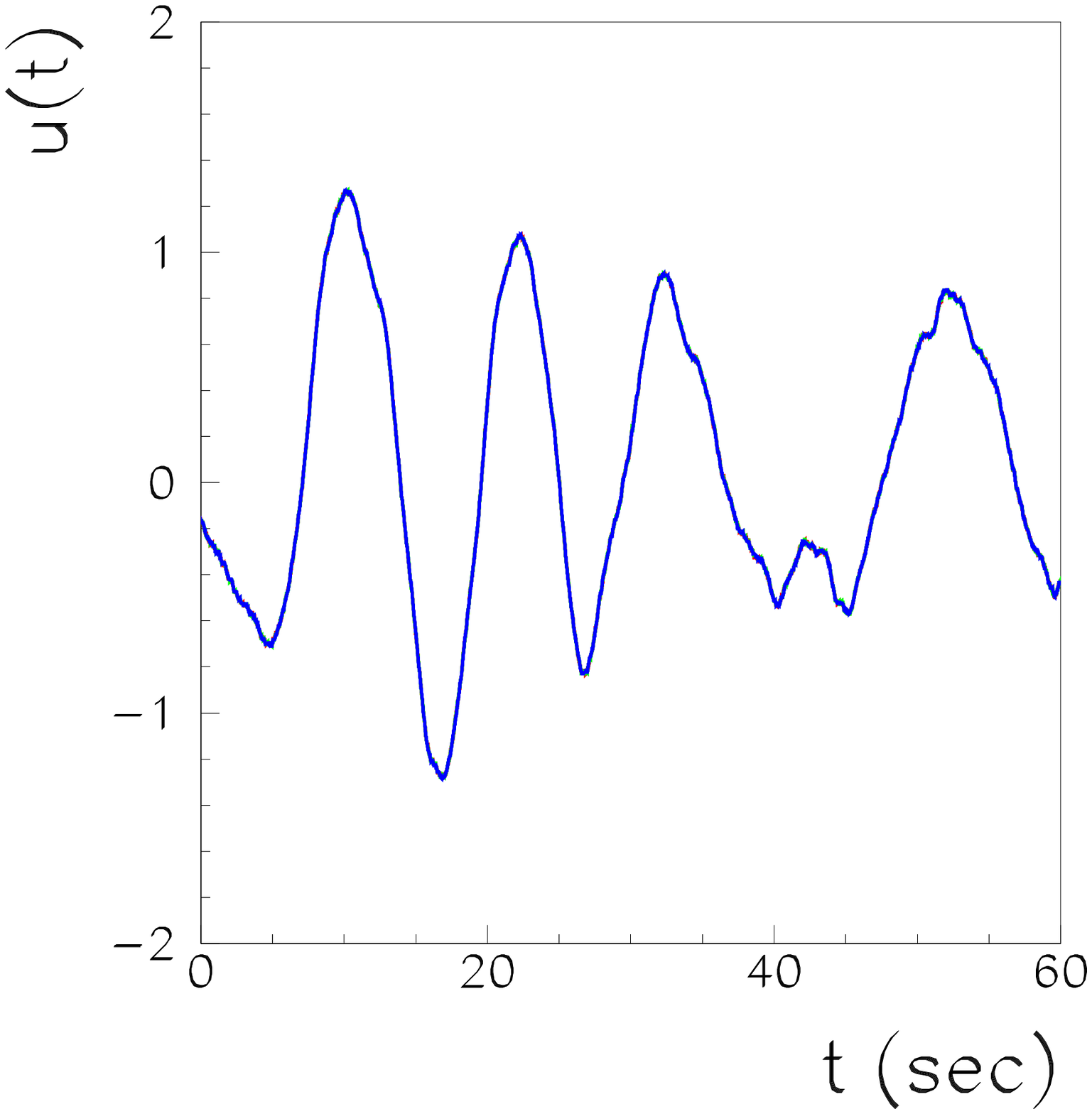} 
&
\epsfysize = 5cm
\epsfbox{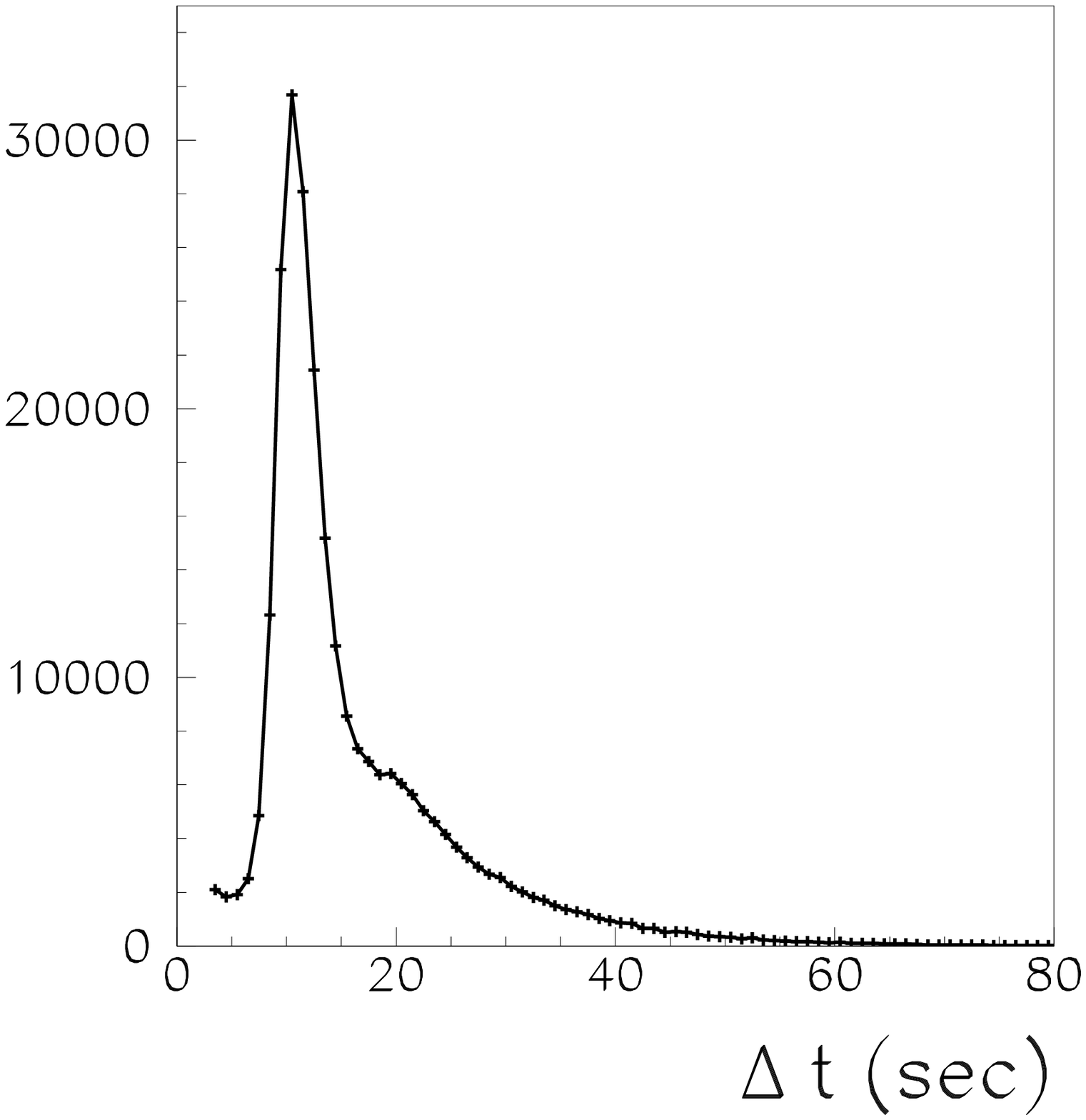}
&
\epsfysize = 5cm
\epsfbox{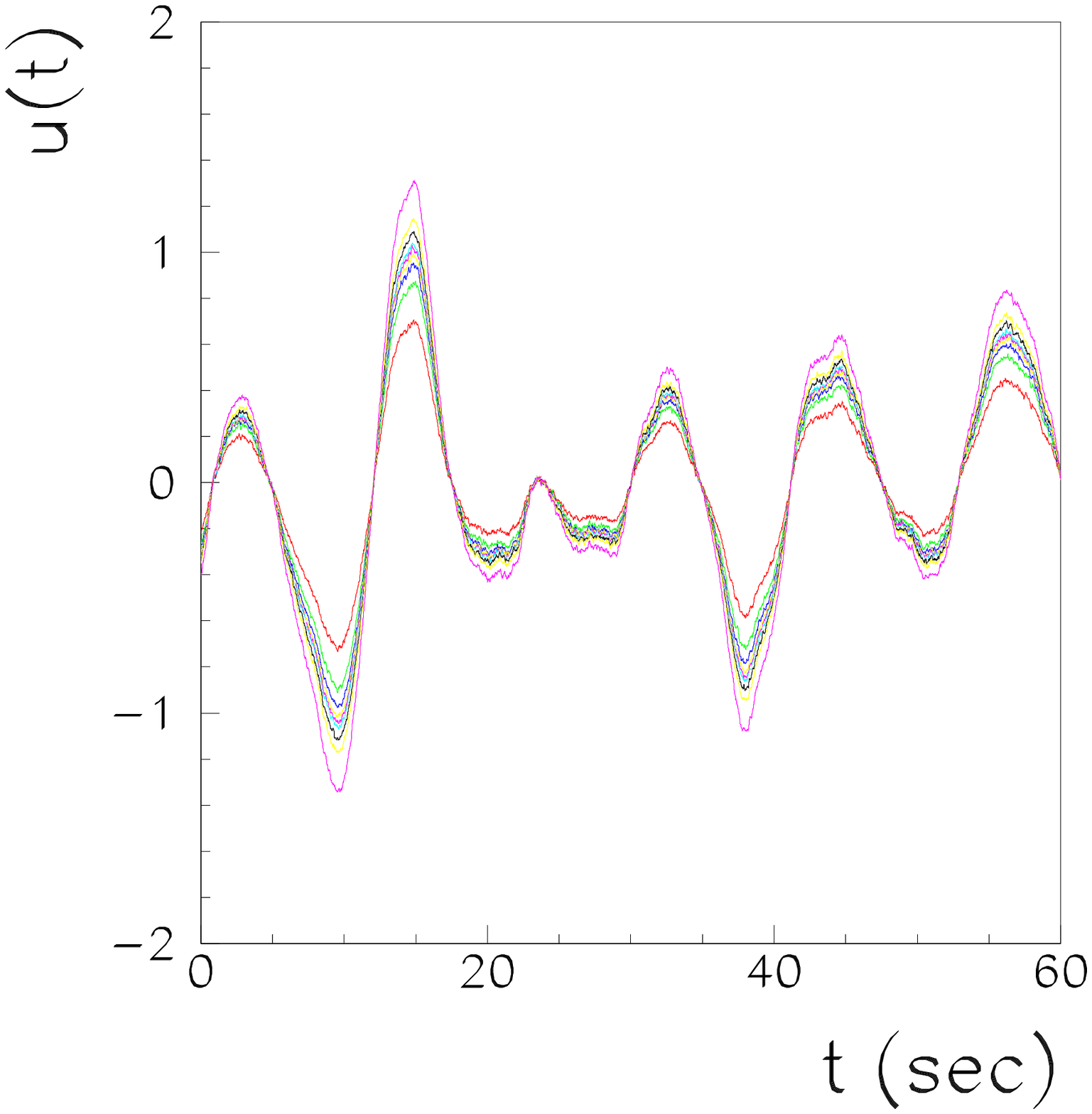}
\end{tabular}
\caption{
Numerical simulation of a noisy nonlinear square model with
structured short-range
connectivity  in regime B, using periodic boundary
condition (eq. \ref{finite}) (a\&b) and open boundary conditions (c). 
N=100 excitatory and N inhibitory units
are placed on a 10x10 square. 
The values of parameters has been chosen in such a way to satisfy
the conditions for the regime B
(specifically,  $\alpha= 50 sec^{-1}$, $N=100$,
 $W_0=h_0= \sqrt{0.25 j_0^2+ 0.25} $, $j_0 = 2(\alpha -0.1)$,
so that $\omega_0= 0.5 rad/sec$).
a: The time behavior of excitatory activity $u_i(t), i=1,\dots,N$ in
the model with periodic boundary condition. (Specifically parameters are
$J_{ij}= \frac{j_0}{8} $ and
$ W_{ij}= W_0/8$  for $|i-j| \mod N = 1,M,M\pm 1,$ and
 $H_{ij}= h_0 \delta(i-j)$, $\Gamma= 0.001$).
b: Histogram of Inter Synchronous Event Intervals of the activity shown in a.
c: The time behavior of excitatory activity $u_i(t), i=1,\dots,8$ in
the model with open boundary condition. Activity is synchronous, but with
different amplitudes. Parameters: 
$J_{ij}= j_0/7.75  
 $ and
$ W_{ij}= W_0/7.75 
$ for $|i-j|= \pm 1, \pm M, \pm (M\pm 1)$, zero otherwise,
  $H_{ij}= h_0 \delta(i-j)$, $\Gamma= 0.04$.}
\label{shortB}
\end{figure}

We perform numerical simulations of the nonlinear network model
with the long-range connectivity 
and
with the short-range connectivity. 
The values of parameters has been chosen in such a way to satisfy
the conditions for the regime B.
Similar results have been obtained in the long-range connectivity
case  (fig. \ref{fig_power_linB}, \ref{figIEI})
 and in the short-range connectivity model (fig. \ref{shortB}).
Noise induces spontaneous oscillations that are synchronous because $\bxi_0$
has real positive elements. 
Fig.\ref{fig_power_linB} shows the synchronous time behavior of the state
 variables $u_i(t)$,
and its power spectrum density, when noise is $\Gamma=0.0004$ in the
N=10 long-range connections model (eq.\ref{infinite}) , in regime B.
The signal $u_i(t)$ looks  aperiodic on long time scale,
being decorrelated over long time scale by the noise.
This spontaneous aperiodic  synchronous  activity
mimic the  spontaneous aperiodic  synchronous activity  observed
under 0.5 mM Ca (and 2.0 mM Ca) concentration in \cite{segev}.
We compute the Inter-Event-Interval (IEI) between two
successive bursting events. A peak of all $u_i$ with intensity above
$0.7$ is defined as a synchronized bursting event. From the sequence $t_n$,
specifying the location of the $n$th event, we compute the IEI histogram 
shown in fig. \ref{figIEI}.
Fig.\ref{shortB}.b shows the IEI histogram of a short-range connectivity
model (eq.\ref{finite}) with N=100 in regime B with noise $\Gamma= 0.001$. 
Notably, both IEI are in a good agreement with the  experimental one
observed  at $0.5$ mM Ca (see fig. 5 of \cite{segev}).
Both experimentally and in the model, the IEI shows a peak 
(at about 10 sec) with a very long tail
(around 40-60 sec the histogram is still significantly non zero).

\begin{figure}[ht]
\vspace*{0.0cm}
\begin{center} \leavevmode 
\begin{tabular}{cccccc}
\hspace{-0.6cm}  a &
\hspace{-0.6cm}
\epsfysize = 4.0cm  
\epsfbox{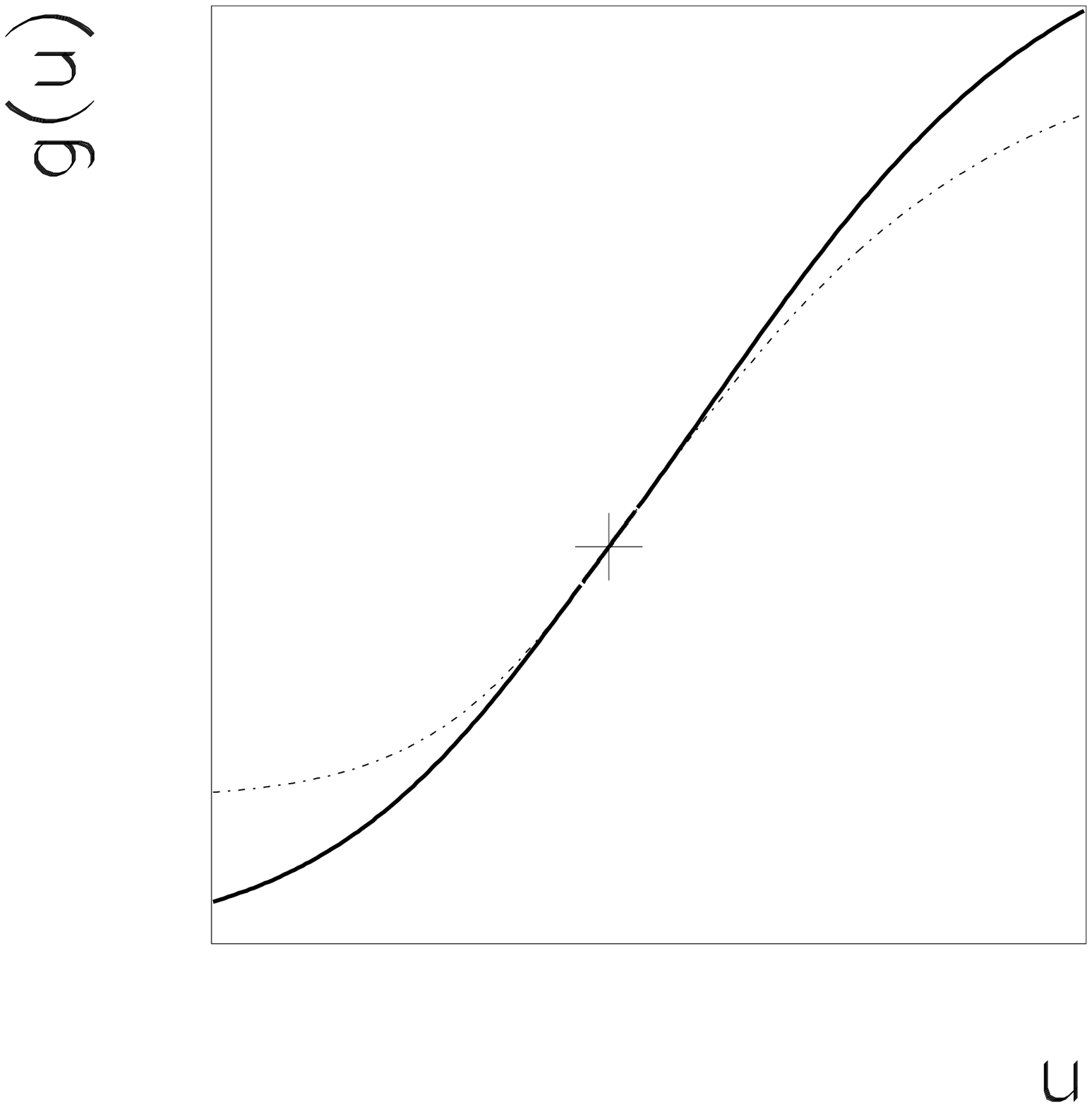}
&
b& 
\epsfysize = 3.9cm 
\epsfbox{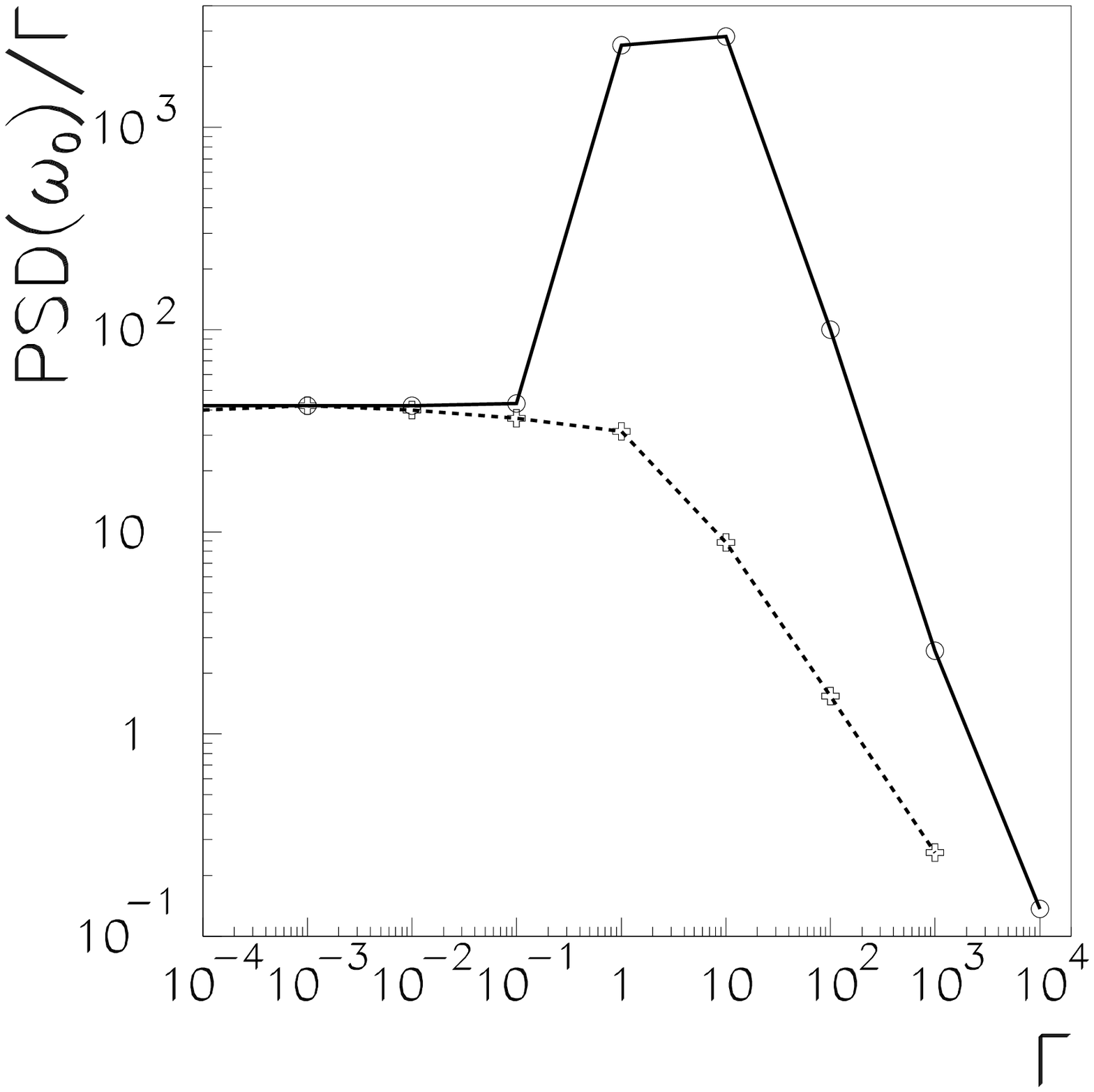}        
&
c &
\epsfysize = 3.9cm 
\epsfbox{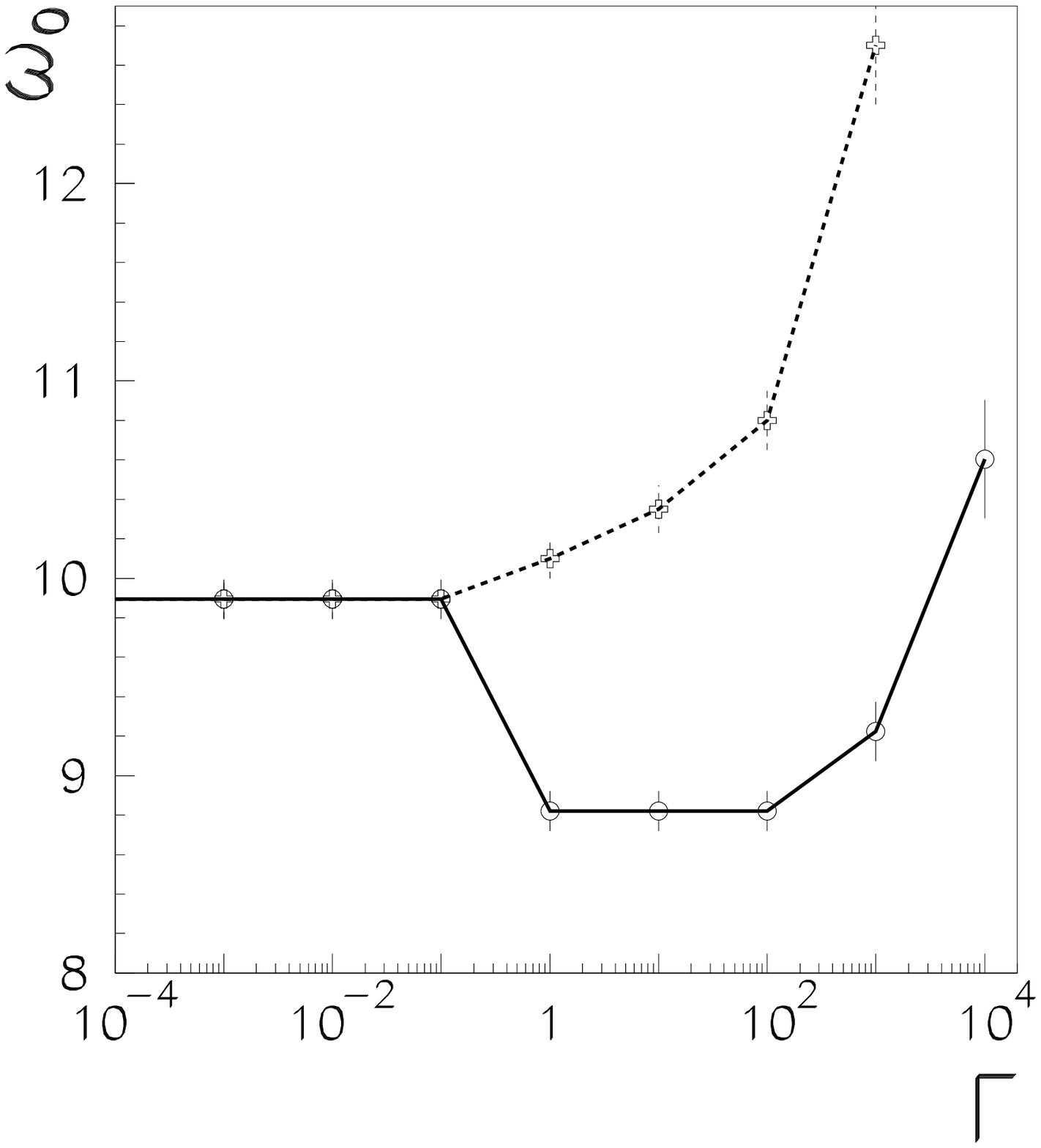}
 \\
\end{tabular} 
\end{center}
\vspace*{0.0cm}  
\caption{
Nonlinearity and stochastic resonance effects.
Numerical simulations of a long-range connectivity model in regime B.
a. Two  activation functions for excitatory
 units $u_i$.  Dashed line shows a saturating activation function
 (class I, used in most of the following simulations),
while solid line shows a class II nonlinear function
whose slope increases before decreasing when
$u$ is raised  from its stationary value (cross).  
b. Ratio $R= PSD(\omega_0)/\Gamma$ as a function of the
noise level $\Gamma$. Dashed line correspond to
the class I activation function shown with dashed line in a, 
while the solid line to the solid line class II activation function. 
We used the following parameters in the simulations:
$N=10$, $\alpha= 50 sec^{-1}$
 $J_{ij}= \frac{j_0}{N} = 9.98 sec^{-1}$,
$W_0=h_0=50.89 sec^{-1}$, so that $\omega_0=10 rad/sec$.
c. The  frequency of the PSD peak
versus the noise level 
(same simulations as in b).
Effect of nonlinearity is evident in both fig. b and c, indeed
for linear system $R$ and $\omega_0$
do not change with noise level.
}
\label{picco}
\end{figure}

Introduction of a  sigmoidal nonlinearity $g(u)$
 does not change  the linear approximation results drastically. 
Numerical simulations show that noise induces similar synchronous
oscillatory activity  both in the nonlinear 
system and in the linearized one (see
Fig.\ref{fig_power_linB}.b). 
  However some classes of  nonlinearity can
show Coherent Stochastic Resonance phenomena \cite{stochasticres}.
We categorize the nonlinearity into
2 general classes (as in \cite{SZJ})
 in terms of how $g_u$ deviates from linearity near the
fixed point $\bar \bu$:
\begin{equation}      \mbox{class I:} ~~~~~~g_u (u_i) \sim u_i - a u_i^3
~~~~~~~~~~ \mbox{class II:}
~~~~~~        g_u (u_i) \sim u_i + a u_i^3 - b u_i^5
\label{gclass}
\end{equation}
where  $a, b >0$,
 and $u_i$  is measured from the fixed point value $\bar u_i$.
Class I and II nonlinearity differ in whether the gain $g'_u$ decreases
or increases (before saturation) as one moves away from the equilibrium
point, and will lead to qualitatively different behavior, as will be shown.
Fig. \ref{picco}.b shows the ratio
$R= \tilde C(\omega_0)/\Gamma$
 between the output power at
the  characteristic frequency $\omega_0$ and
the power of the noise $\Gamma$,  versus the noise level,
using the two nonlinearities shown in fig.\ref{picco}.a
(dashed line for class I ,  solid line for class II), in a long-range
connectivity model.
Class II nonlinearity
can enhance the ratio between the height of the PSD peak and the
strength of noise, for a critical range of noise intensities. 
Solid line in Fig. \ref{picco}.b 
 shows the typical maximum that
has become the fingerprint of {\it Stochastic Resonance} phenomena
\cite{stochasticres}.
Fig. \ref{picco}.c shows that the oscillation frequency changes (slowly) with
the noise level $\Gamma$.
%

%


\section{Regime C Dynamics}
\label{secC}

If the excitatory-to-excitatory connections are stronger, so that
$\Re[\lambda_{n=0}] = \alpha- j_0/2 \geq 0$, but not too strong, so
that it is still $ \Im [\lambda_0]
\neq0 $,        
then spontaneous oscillatory  activity arises also without  noise.
In particular, spontaneous 
periodic oscillations arise in the
linear approximation  if $\Re[\lambda_0]=0 $ and $\Im[\lambda_0]\neq 0$, this
is a critical point separating the regimes  $\Re[\lambda_0]<0 $,
$\Im[\lambda_0]\neq 0$  (regime B) and $\Re[\lambda_0]>0$,
$\Im[\lambda_0]\neq 0$  (regime C).

In the regime C linear analysis predict synchronous periodic activity with
diverging amplitude, that becomes stable when nonlinearity is taken into
account.
 We focus on class I nonlinearity (eq. \ref{gclass}),
 when  $g_u$ deviates from linearity near the
fixed point $\bar \bu$ due to saturation.
 This case includes most standard
sigmoids used in modeling, including logistic sigmoids, hyperbolic tangent
 and rounded
threshold-linear models with a soft saturation at high input level.
Equation (\ref {ddu}) then becomes
\begin{equation}
[(\alpha + \partial_t )^2]{\bu}  -
[ (\partial_t +\alpha)  {\sf J}
- {\sf H} {\sf W} ]  g_u (\bu ) =  - H F(t)
\label{eqnsgu}
\end{equation}
where  by $g_u(\bu)$ we
mean a vector with components $[g_u(\bu)]_i = g_u(u_i)$.

\begin{figure}[ht]
\vspace*{-0.1cm}
\begin{center} \leavevmode 
\begin{tabular}{cccc}
a & b & c & d\\                                                              
\epsfysize = 4.4cm 
\epsfbox{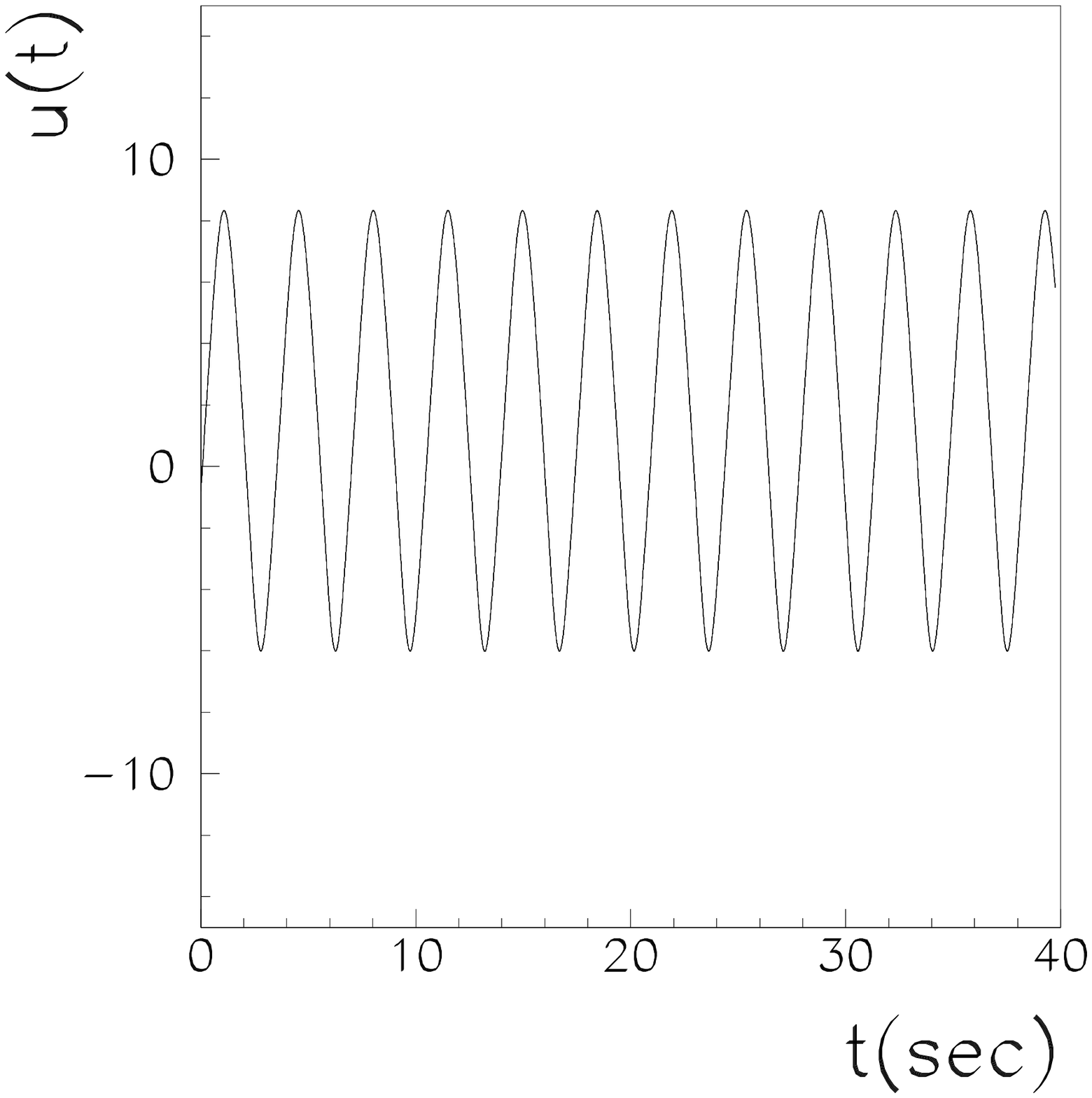} 
&
\epsfysize = 4.4cm 
\epsfbox{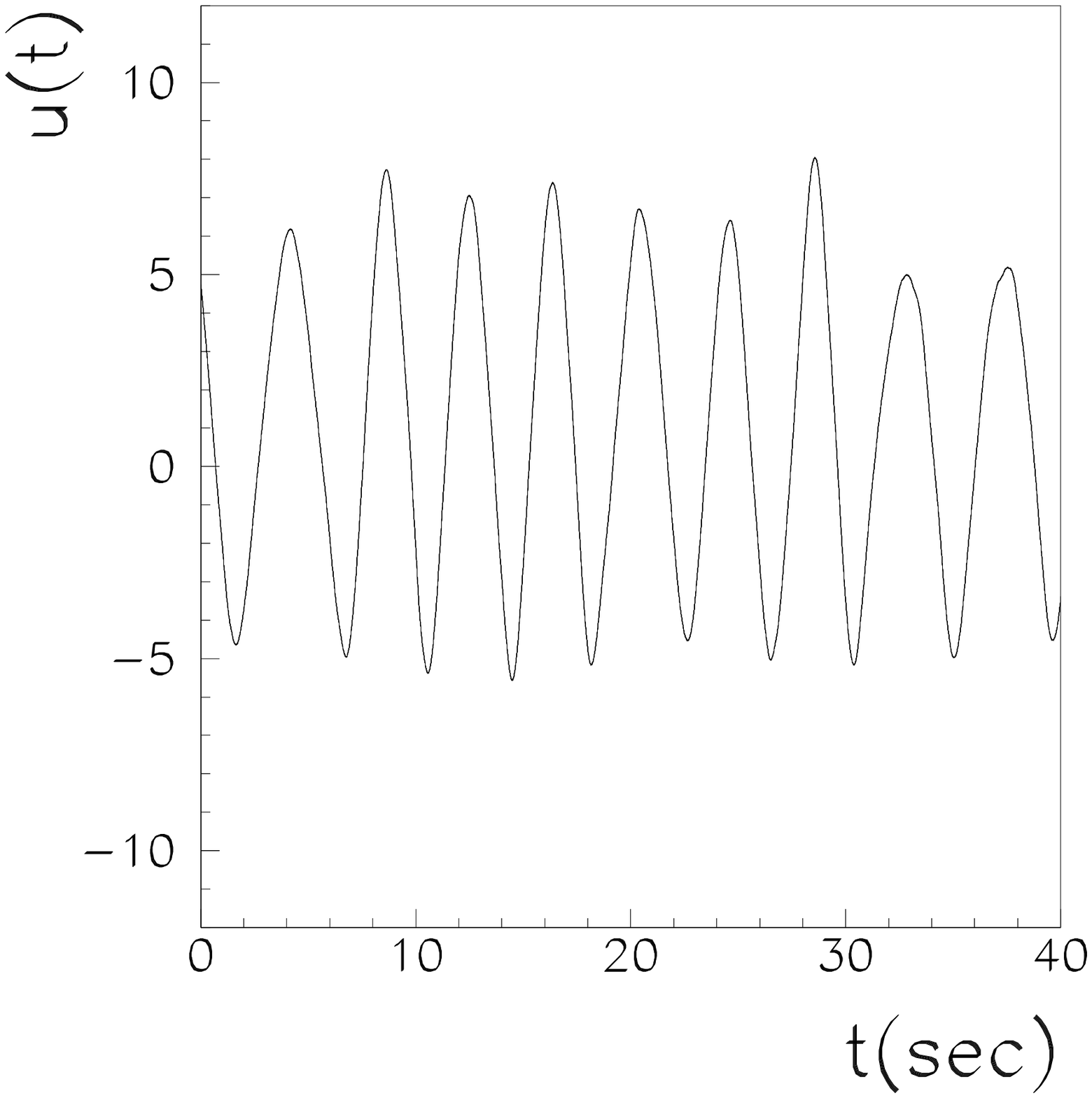}     
&
\epsfysize = 4.4cm 
\epsfbox{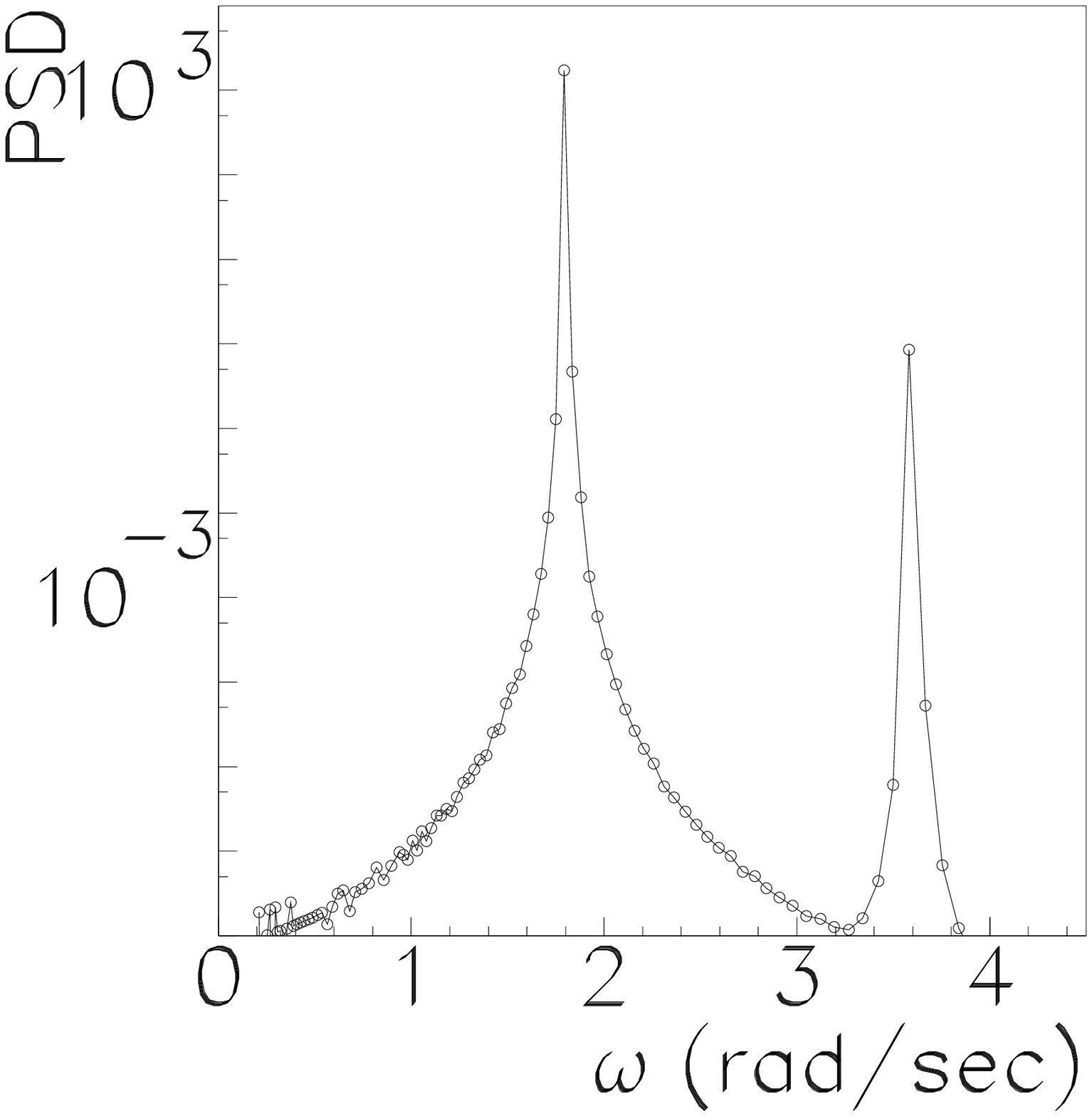}
&
\epsfysize = 4.4cm
\epsfbox{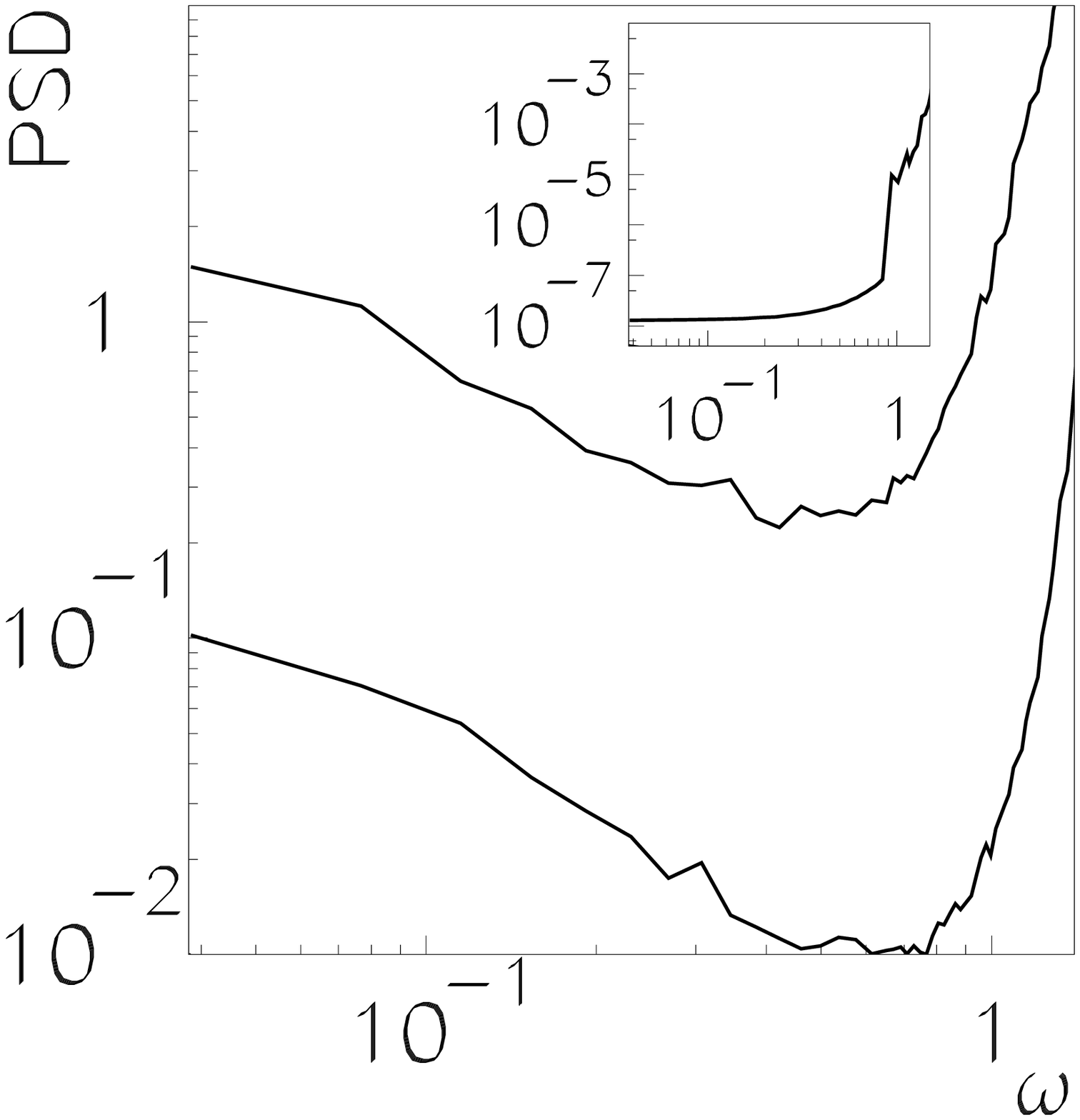}
\end{tabular}  
\end{center}
\vspace*{-0.1cm}
\caption{
Regime C simulations of the nonlinear model
with long-range connectivity (eq.\ref{infinite}).
We use the following parameters
$N=10$, $J_{ij}= \frac{j_0}{N} = \frac{ 2(\alpha +0.07)}{N} $,
and $W_0=h_0= \sqrt{0.25 j_0^2+0.25} $.
The activation function 
for the excitatory unit is shows in Fig.\ref{fig_power_linB}.a.
a,b: Time behavior of the state variables $u_i(t)$, 
in the  numerical simulation of the  
isolate EI  long-range connectivity network in regime C. 
(Lines $u_i(t)$, $i=1,..,10$,  overlaps each other because of synchrony).
a: regime C, noiseless nonlinear system.
Synchronous periodic oscillatory activity is shown.
b: regime C, noisy nonlinear system with  $\Gamma= 0.01$. 
The activity is almost periodic,
but noise  decorrelates the signal over long time scales.
c:
Power Spectrum Density of the  activity shown in a (lin-log scale).
 The peaks in the first and second harmonic mark clearly
the collective periodic activity.
d:Log-log plot of the  Power Spectrum Density
 of the  activity in regime C at $\Gamma= 0.01$ (upper curve),
$\Gamma= 0.001$ (lower curve) and $\Gamma=0$ (inset).
Noise induces a broad peak at low frequency. 
}
\label{fig_u_aaa}
\end{figure}

Fig. \ref{fig_u_aaa} shows the simulation results of a long-range connectivity
 model in regime C,
with and without noise.
 Fig.\ref{fig_power_linB}.a shows the saturating function that
we have used in our numerical nonlinear simulations for excitatory
units.  
Simulation results of N=100 short-range connectivity model with
M=L=10 in regime C, shown in Fig.\ref{shortC},
 are in qualitative agreement with the
long-range connectivity model results.
As shown in Fig. \ref{fig_u_aaa}.a and Fig.\ref{shortC}.a
the noiseless nonlinear numerical simulations
shows stable  synchronous  periodic oscillatory activity $\bu$. 

%
%
%
%

We check numerically that
this spontaneous periodic oscillations behavior
 holds only 
for a particular
range of parameters (regime C).
For example, starting from the regime C parameters used in
Fig.\ref{fig_u_aaa},
 for lower excitatory connections ($j_0=99.86 < 2\alpha$)
we get regime B aperiodic oscillations, while 
increasing j0 beyond the regime C range
(e.g., at $j_0=103 > \sqrt{h_0 w_0} $) 
 the network jumps to a new stable fixed point
(and it starts to oscillate
around the new  fixed point if there's noise).
%

\begin{figure}[ht]
\vspace*{-0.1cm}
\begin{center} 
\epsfysize = 6cm
\epsfbox{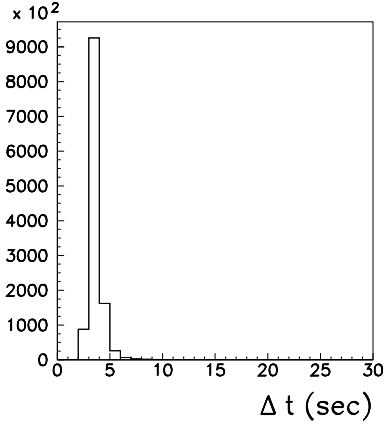}
\end{center}
\caption{Histogram of Inter Synchronous Event Intervals, of
the activity shown in figure \ref{fig_u_aaa}.b (nonlinear model
with long-range connectivity 
in regime C with $\Gamma= 0.01$ )}
\label{figIEI2}
\end{figure}

\begin{figure}[ht]
\vspace*{-0.1cm}
\begin{center} \leavevmode 
\begin{tabular}{ccc}
a & b & c\\                                                              
\epsfysize = 4.4cm 
\epsfbox{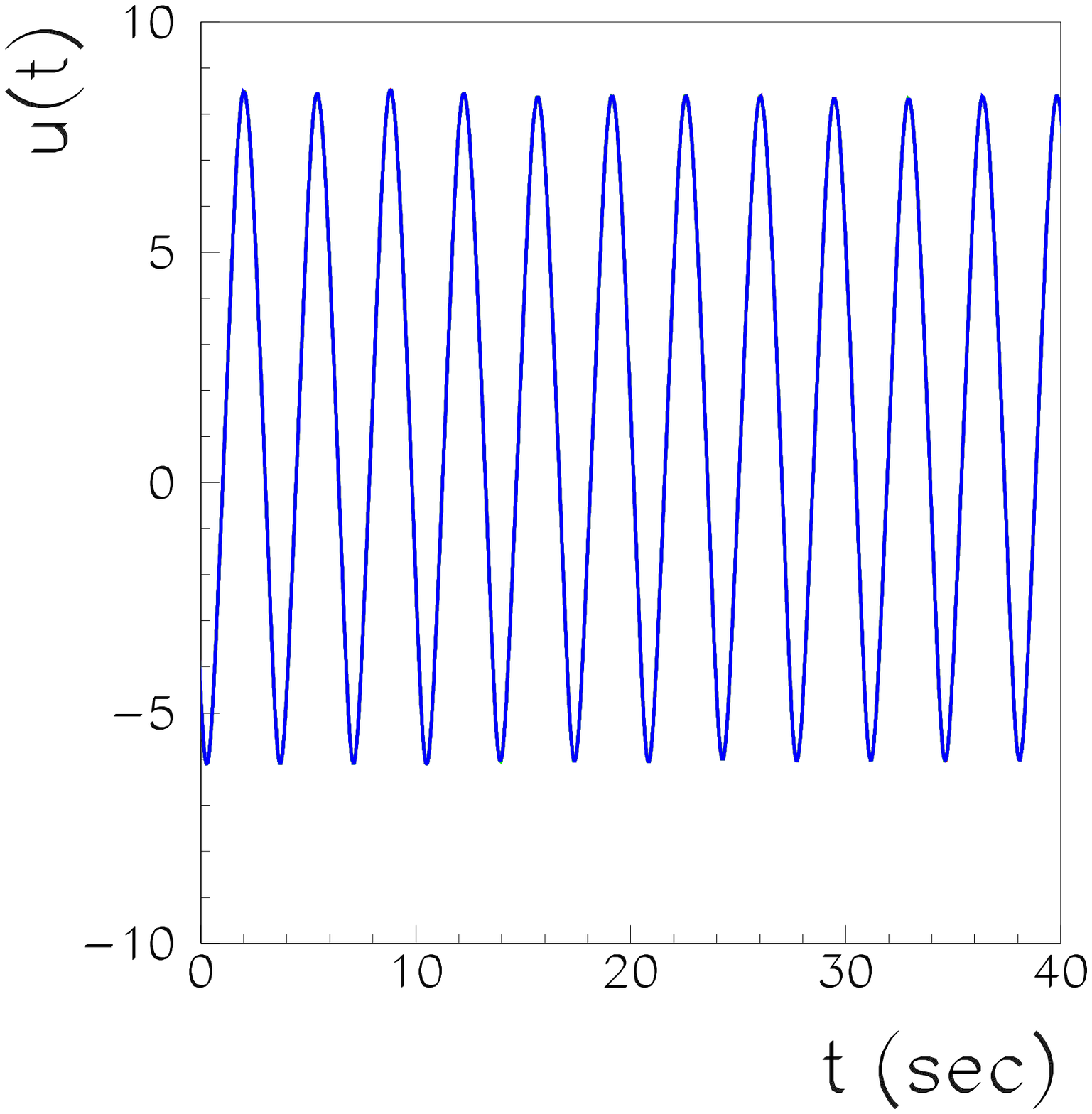} 
&
\epsfysize = 4.4cm 
\epsfbox{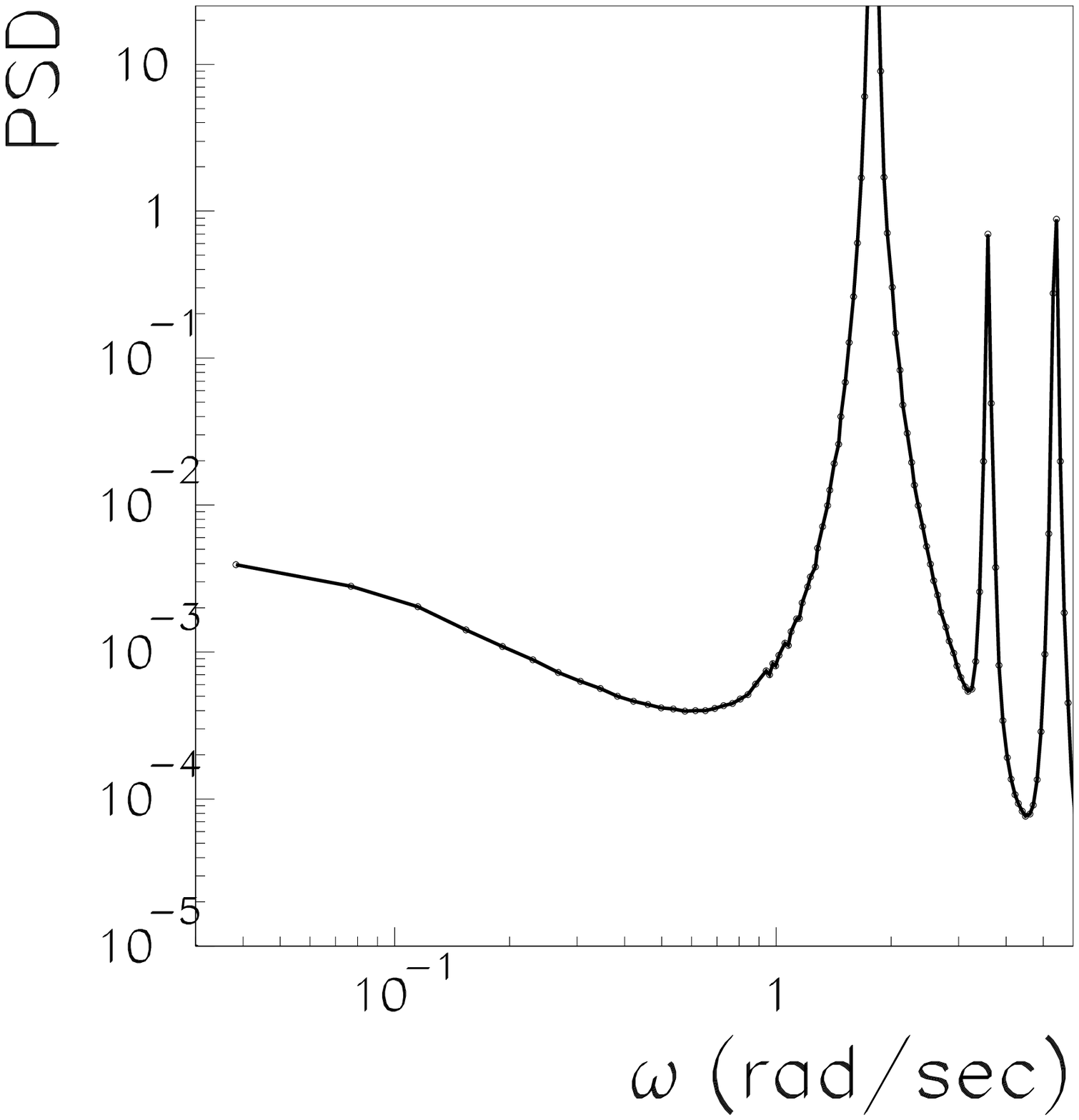}     
&
\epsfysize = 4.4cm
\epsfbox{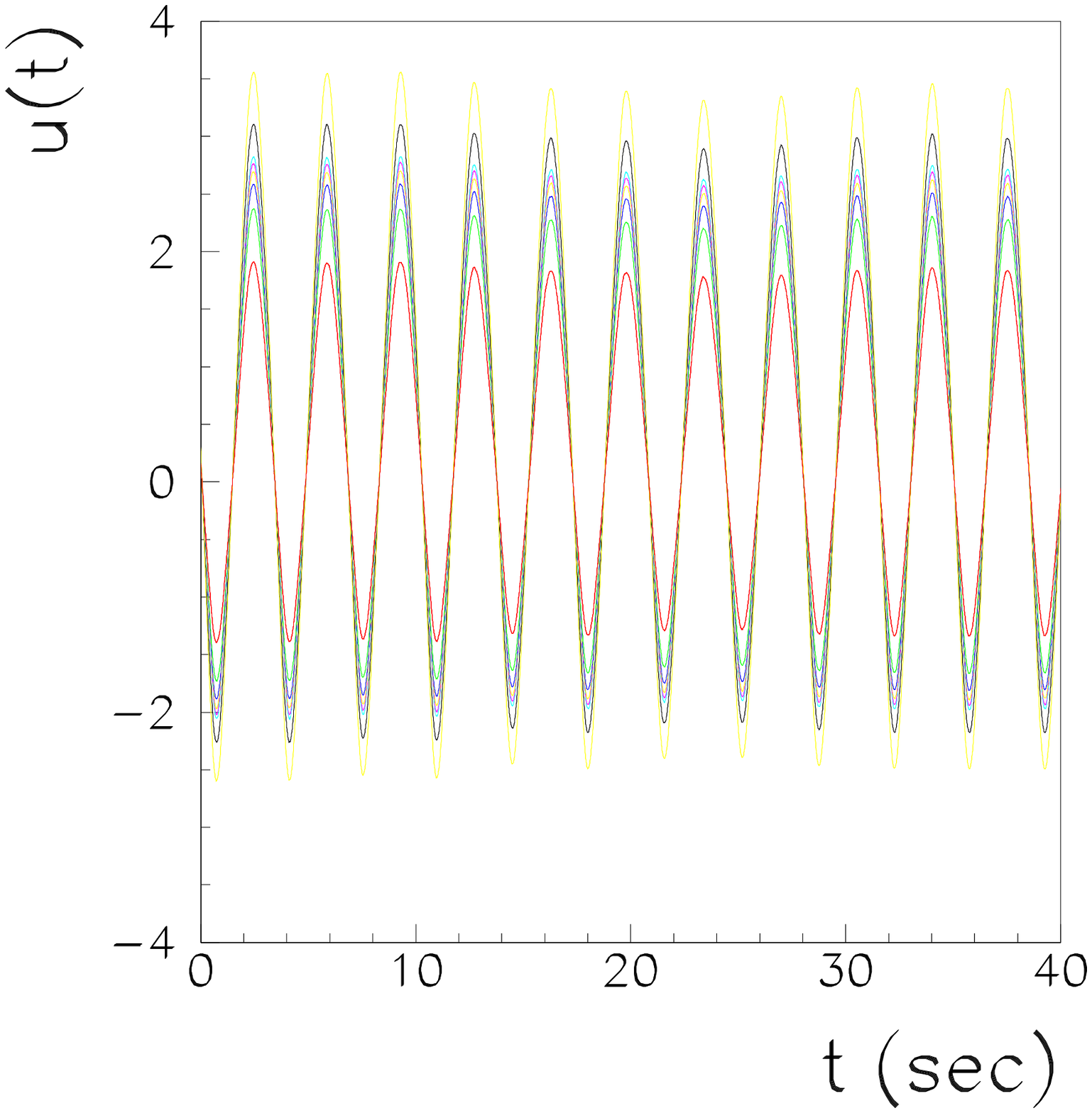}
\end{tabular}  
\end{center}
\vspace*{-0.1cm}
\caption{
Numerical simulations of nonlinear model with short-range connectivity
(\ref{finite}) in regime C, using periodic boundary condition (a\&b) and
open boundary condition (c).
N=100 excitatory and N inhibitory units
are placed on a 10x10 square. 
The values of parameters has been chosen in such a way to satisfy
the conditions for the regime C (specifically,
$N=100$, ${j_0} ={ 2(\alpha +0.07)} $,
$W_0=h_0= \sqrt{0.25 j_0^2+0.25} $).
a:Time behavior of the state variables $u_i(t)$ (synchrony) 
in the periodic boundary condition connectivity case (\ref{finite}).
Specifically parameters are: $J_{ij}= j_0/8$, $W_{ij}=W_0/8$,
 for $|i-j| \mod N =  1, M,  M\pm 1 $, $H_{ij}= h_0 \delta(i-j)$ )
b: PSD of the activity $u_i(t)$ shown in a.
c: Time behavior of the state variable $u_i(t)$ for $i=1 \ldots 10$ 
in the open boundary condition model. Activities 
of the units are synchronous but with
different amplitudes. (Specifically parameters are $J_{ij}= j_0/7.75$, $W_{ij}=W_0/7.75$,
 for $|i-j| = \pm 1, \pm M, \pm M\pm 1 $, $H_{ij}= h_0 \delta(i-j)$. )
}
\label{shortC}
\end{figure}

 As shown in Fig.\ref{fig_u_aaa}.c and Fig.\ref{shortC}.b
 the PSD in regime C has two high peaks
at  first and second harmonic of the periodic network activity,
 resembling the experimental results of \cite{segev}. 
Moreover \cite{segevl} has put in evidence
 experimentally
 the existence of a broad peak at low frequencies in the PSD,
indicating long time positive correlations. 
Our model reproduces these results, indeed
numerical simulations of  regime C  in presence of noise, with
both type of connectivities,
 show a broad peak at low frequency in the PSD (see fig.\ref{fig_u_aaa}.d).
 The broad peak is absent when the
noise is absent. 
The low frequency's behavior  indicates a long time positive
autocorrelation in the activity of the network.
 This behavior cannot be
accounted by the IF model of Segev et al. (see fig. 3.c in \cite{segevl}). 
In order to compare the IEI behavior of our model in the two regimes, the
 IEI histogram for the noisy model in regime C has been computed (see fig.
\ref{figIEI2}).
It looks totally different
from the IEI histogram in regime B. 
In the regime C, for both the long-range and short-range connectivity
cases,
the long time tail of the IEI disappears. 
There is only a narrow high peak corresponding to the period of the signal.
Therefore, one prediction of our model is that the
experimental IEI at 1mM Ca concentration
should 
 have a high peak (periodicity) with a fast decay similarly to
fig. 5
 (differently then at 0.5 mM Ca).

\subsection{Fading of periodicity at long times}

Another effect, put in evidence experimentally \cite{segev},
is the decrease in time of the periodicity of the bursting activity 
under 1 mM Ca on the time scale of several minutes.
According to Segev et al \cite{segev}, this effect
could be related with the possibility of network adaptation, that still
need to be verified experimentally.
 
In our model we identify two parametric changes
which result in a decrease of the periodicity,
namely the increase of noise level 
 and  the depression of excitatory connections.
The depression of the synapses, 
 during the experiments, could be justified in the light
of adaptation processes: shortly after 1 mM Ca is added the
excitatory connections increases their efficacy, while
several minutes after 
the system possibly  adapt to the new concentration and the
efficacy of connections decreases to the original values.  
Also an increase of the noise during time cannot be 
excluded in principle. Even though all conditions,
 like temperature and humidity, 
are kept constant during the experiment, an increase of the
noise of the system is possible.

So, according to our analysis,
 adaptation (which is related to the depression of
the excitatory connections) is not the only
 possible explanation of the fading of periodicity observed
experimentally, since simply an increase of the level of 
intrinsic noise 
is able to account for the observed  fading of periodicity.

\begin{figure}[ht]
\vspace*{0.0cm}
\begin{center} \leavevmode 
\begin{tabular}{cc}
a & b 
\\
\hspace{-0.6cm}
\epsfysize = 5.2cm 
\epsfbox{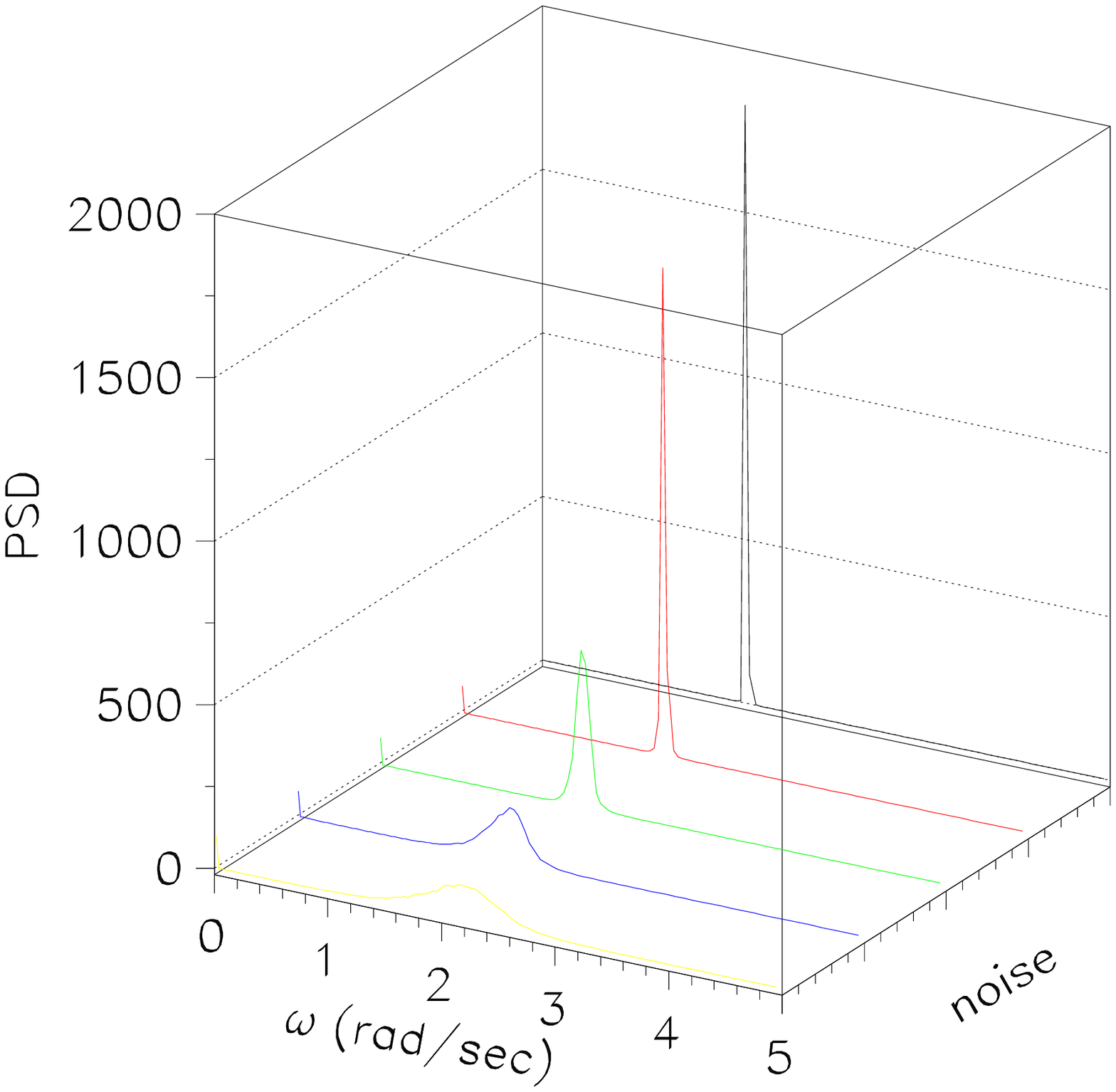} 
&
\hspace{-0.6cm}
 \epsfysize = 4.2cm
\epsfbox{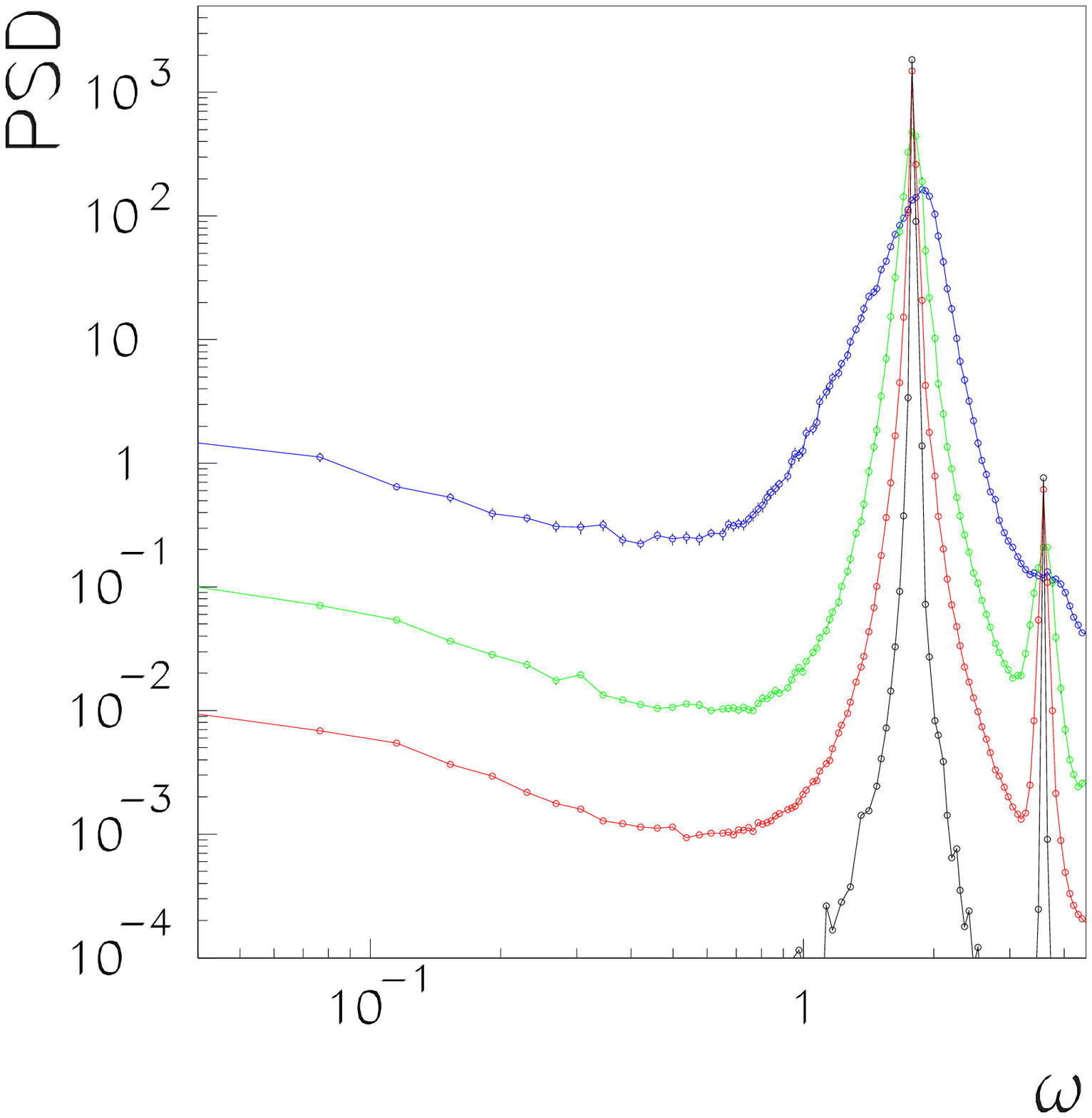} 
\\
\end{tabular} 
\end{center}
\vspace*{0.0cm}
\caption{
a:
Power Spectrum Density of the excitatory units activity
of the noisy nonlinear networks with long-range connectivity 
in regime C, plotted 
at different noise levels.         
 Black line: no-noise $\Gamma=0.0$, 
red line: $\Gamma=0.0001$ ,
green line: $\Gamma=0.001$ ,
 blue line: $\Gamma=0.01$ ,
 yellow line: $\Gamma=0.1$.
Noise make the high periodic activity fades away.
b: Log-log plot of the  Power Spectrum Densities shown in a.       
The two high peaks in the PSD correspond to the first and second harmonics
of the activity period.  
The low frequency power spectrum distribution shows 
a broad peak in $\omega=0$.
}
\label{fig_u_aaa_noise}
\end{figure}
\begin{figure}[ht]
\vspace*{0.0cm}
\begin{center} \leavevmode 
\begin{tabular}{cc}
a & b 
\\
\hspace{-0.6cm}
\epsfysize = 5.2cm 
\epsfbox{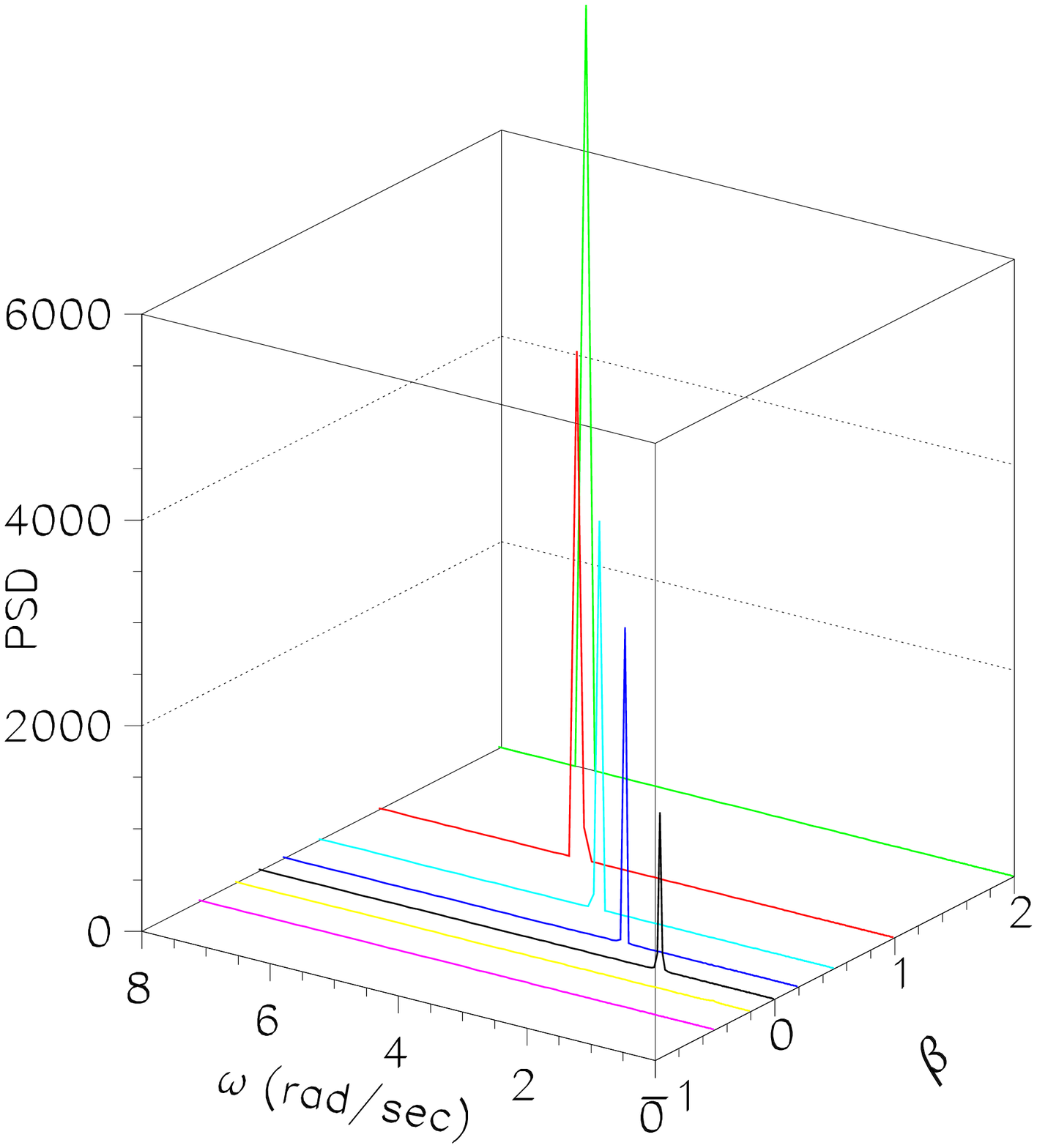} 
&
\hspace{-0.6cm}
 \epsfysize = 4.2cm
\epsfbox{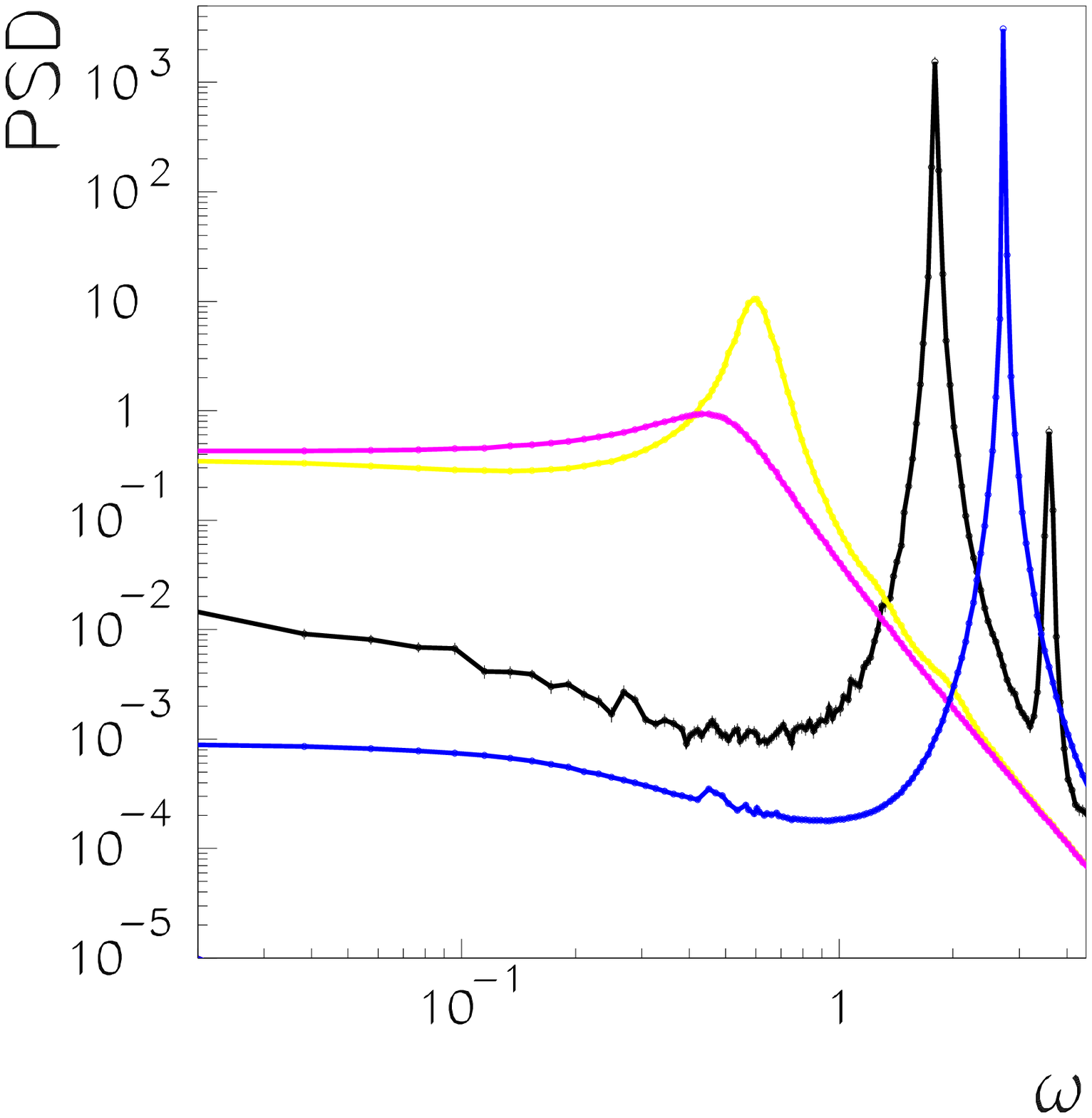}
\\
\end{tabular} 
\end{center}
\vspace*{0.0cm}
\caption{
a:
Power Spectrum Density of the excitatory units activity
of the noisy nonlinear long range networks,
 at $\Gamma=0.0001$, plotted 
versus $\beta$, $j_0=\beta+\bar j_0$, $W_0=\beta+\bar W_0$
(where $\bar j_0 = 2(\alpha +0.07) $ and 
$\bar W_0= \sqrt{0.25 j_0^2+0.25} $ correspond to regime C).
  Increasing $\beta$ the excitatory connections become
stronger.  
Decrease of $\beta$ (i.e. decrease of excitation) make the 
high periodic activity fades away.
b: Log-log plot of some of the Power Spectrum Densities shown in a.        
 Blue line: $\beta=0.2$
back line: $\beta=0$, 
yellow line: $\beta=-0,2$,
 magenta line: $\beta=-0.5$.
The two high peaks in the PSD correspond to the first and second harmonics
of the activity period.  
The low frequency power spectrum distribution shows 
a broad peak in $\omega=0$ that decreases when excitation ($\beta$)
 decreases, fading away.
}
\label{aaa_LTD}
\end{figure}

To show this,
 we have computed numerically the PSD for different
values of the noise and for different
values of the  excitation. 
The results are reported
in fig. \ref{fig_u_aaa_noise} and fig. \ref{aaa_LTD} respectively.
In both the figures it is evident that a  fading of periodicity
(i.e., a decrease of the first and second harmonic
peaks) occurs both when noise increases and when excitatory connections
are depressed.
Looking at 
Fig. \ref{fig_u_aaa_noise} and \ref{aaa_LTD} we see that
the two parametric changes  
(noise-change or excitatory connections decrease)
can be discriminated since their effects on the PSD are different 
at  low frequency.
In Fig. \ref{fig_u_aaa_noise} we note that an increase of
the level of noise leads to 
 an increase of the broad peak at $\omega=0$.
Instead in Fig. \ref{aaa_LTD} it is evident that a depression of
the excitatory connections eigenvalues $j_0,W_0$ 
induces a decrease of both the high peaks and
the broad peak at $\omega=0$.

However, the experimental results at present do not allow to discriminate between the
two possible explanations because, as far as we know,
there are not measurement of the low frequency PSD after the fading of
periodicity occurring on the time scale of several minutes. 
 It could also be that both the phenomena occur and
contribute to the fading of periodicity.

An indication that noise may changes during time
comes from  the  work of T. Tateno \cite{pre02}
 showing the PSD of the activity of a rat cortical network after 12,18, and 58
 days in vitro (DIV). 
While PSD at 12 DIV seems to correspond
to a noiseless or low-noise system,
 the figures at 18 and 58 DIV brings to mind the possibility that the
noise is increased (in particular a low frequency peak becomes visible, 
even though the lin-log scale doesn't highlight it).
A strong decrease of periodic activity occurs as numbers of days increases.

\subsection{Phase-locking between excitatory and inhibitory
population}
Besides the periodic synchronous bursting activity analyzed 
until now, our model can exhibit periodic but not synchronous
bursting activity. This appear when there's a phase lock between neurons
involved in the burst. Mathematically it is described by
%
$u_i(t) \propto |\xi_i| \cos (\omega_0 t - \phi_i)$, 
  We can describe
both synchronous and phase-locked periodic activities cases
 by writing $u_i(t) = \xi_i \e^{-\i \omega_0 t} + \cc$, taking
the $\xi_i$ real in the first case and complex ($\xi_i = |\xi_i|\e^{\i
\phi_i}$) in the second.  

\begin{figure}[t]
\vspace*{0.0cm}
\begin{center} \leavevmode \epsfysize = 5.5cm 
\epsfbox{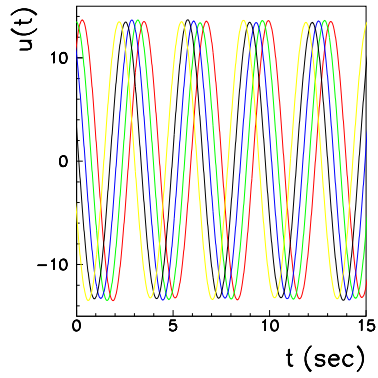} 
\end{center}
\caption{
Phase-lock oscillatory activity $u_i(t)$, $i=1,\ldots,5$.
N=100 excitatory and N inhibitory units
are placed on a 10x10 square. $J_{ij}$ and $W_{ij}$ are 
positive when
  $|i-j| \mod N= 1,M,M\pm1 $
and are negative  when $|i-j| \mod N = 3, 3M, 3M \pm1$, 
all the other elements are null. $H_{ij} = h_0 \delta_{ij}$ as usual. 
The values of parameters has been chosen in such a way to satisfy
the conditions for the regime C.
 }
\label{m}
\end{figure}

A phase locked oscillation arises when the dominant eigenvector is complex.
Fig. \ref{m} shows spontaneous  oscillatory activity in
regime C with a phase shift $\phi_i = 20 \pi/N  i $,  
between all the units $u_i$ involved in the oscillation.  
This is achieved using both positive and negative
connections.
Negative connections can be simply implemented via inhibitory interneurons
with very short time membrane time constants.

In all simulations of previous sections the excitatory activity is synchronous,
while the excitatory population and inhibitory population periodic 
activity are phase locked,  
and the phase shift between the two ensemble
depends from network parameters. 
In all the simulations  of the previous figures
 the  $u(t)$ {\it vs.} $ v(t)$ 
phase shift is  about few thousandths of a period,
in good agreement with the theoretical result
 $arctg(\omega_0/\alpha)$ coming from  the linear approximation
 analysis of our model.
In the experiments of Segev et al.\cite{segevl,segev}
all the oscillatory population (both excitatory and inhibitory)
appear synchronous.
Indeed the predicted phase-shift is so tiny that  cannot be
clearly distinguished from zero.

\section{Summary and Conclusions }
\label{conclusion}

In the present paper we analyzed the effect of uncorrelated white noise
on the spontaneous dynamics of the EI model.
Firstly we study the evolution of the system in the linear approximation 
and put in evidence the existence of two different regimes, B and C.
The transition from B to C regime is induced by increasing the strength of
the excitatory synapses. Then nonlinear corrections have been introduced
numerically. The noise induces in the two regimes very different behaviors.
In the regime B the presence of noise induces collective synchronous
aperiodic oscillations in the neural activity.
In this regime nonlinearity does not change the synchronous oscillatory
activity but can induce stochastic resonance.

In the regime C stable spontaneous periodic oscillations can arise only in 
presence of nonlinearity.  The oscillations are present also in noiseless
conditions. The noise produces a broad peak at low frequency and a fading of
periodicity at high level of noise.

%
The stochastic  model presented is able to account for 
the experimental results of Segev et al.
In particular it accounts for both the observed synchronous periodic regime 
(at 1.0 mM Ca concentration - regime C ) and aperiodic regime (at 0.5 mM and 2.0 mM Ca
concentration - regime B), and transition from one regime to another.
It accounts for the long time positive correlations of the bursting
activity, the IEI and spectral features of the activity, 
and the observed fading of periodicity. 

All to all connection matrices whose elements
are rescaled with N is a very  simplified model that, even though useful
for analytical calculations,
is not realistic. Recent
 estimation of connectivity in in-vitro rat 
cortical networks has shown  arborization of the neurons
of $1.2\pm 0.5 mm^2$ \cite{pre02}. 
A more plausible connection matrix for our model therefore is a structured 
finite range
connection structure. 
We have simulated a network of 100 excitatory and 100 inhibitory
units: each couple of excitatory and inhibitory units is placed 
in a square  lattice. Apart from the units on the boundary, 
each excitatory unit is connected to
its four nearest neighbors and to its four next-nearest neighbors.
Inhibitory-to-excitatory connections H are still local, with each inhibitory
unit connected to only one excitatory unit. 

In this structured short-range connections framework we can also account 
for the  size-independence of the activity  networks.
Indeed Segev et al. \cite{segevl} observe a similarity in the activity of
different-size networks. In order to account for this invariance,
they suggested that the networks have a 
self-regulation process that can be achieved, for example, by an adjustments 
of neural efficacies or
neuronal firing threshold.
However, the similarity of activity find a simple explanation 
in the scenario where each unit is connected  only to other units that are
within a critical radius.
%
While in a all-to-all connection model, in order to keep the activity
constant the strength of the connections should scale with the size of the
model,  in a short-range connection model the strength of connections 
should not scale with the network's size.  
In order to scale the strength of connections with the size of the network
a sort of self-regulations should be invoked.
In the short-range model we get the scale-invariant activity as a bonus.
Indeed the strength of the principal eigenvalue depends on
the number of connections for each unit in average,
  and not on the total number of units.


According to our model,  networks with one size too small with respect to 
the range of connections (such that boundary effects becomes relevant
and affect the average number of connections for each unit) 
shows  dissimilarity from the larger networks activity,
 at the same environment
conditions (same Ca concentration, density, etc).

Finally it is worth to mention two points.

i) The occurrence of different regimes of activity in in-vitro  cortical
networks has been also shown in the recent work of Tateno et al.
\cite{pre02}. Fig. 6.a of \cite{pre02} shows the PSD that we recognize as
typical of regime C, while for example we recognize indications of a
possible broad peak at low frequency in the PSD shown in fig. 6.b and 6.c.   

ii) The "anomalous" low frequency broad peak in the PSD of neural activity 
is not specific to cortical in-vitro networks, it has been 
observed in vivo, for example, in the pulse trains of nerve cells belonging to
various brain structures (such as auditory nerve \cite{2u} and the 
mesencephalic reticular formation \cite{3u}) and in IF models \cite{usher}. 
In \cite{usher} it has been related to the metastability 
of high activity patterns in the
presence of noise (patterns that diffuse throughout the system). 
The "anomalous" peak has been also pointed
out recently in the activity of the Suprachiasmatic
nucleus \cite{fano}.
 It correspond to an "anomalous" behavior of the Fano factor at large
times.  
We claim that  in the Suprachiasmatic Nucleus
neurons, as in the rat cortical cultures,
the low frequency peak in the PSD is just the result of the
 interplay between nonlinearity and the intrinsic noise.

%
%
%
%
%
%

%
%
%
%
%
%

\vspace{-0.4cm}
%
%
%
%
%
%

\end{document}